\documentclass[apj,twocolumn,chicago]{emulateapj}
\usepackage{amsmath}
\usepackage{float}
\usepackage{nicefrac}
\usepackage{rotating}
\usepackage{rotfloat}
\usepackage[none]{hyphenat}
\usepackage{enumitem}
\usepackage{afterpage}
\usepackage{datetime}
\usepackage{color}
\usepackage{url}
\usepackage[flushleft]{threeparttable}
\usepackage{xspace}

\usepackage[colorlinks]{hyperref}
\usepackage{etoolbox}

\hypersetup{breaklinks=true,colorlinks=true,linkcolor=blue,citecolor=blue,filecolor=blue,urlcolor=blue}
\makeatletter
\patchcmd{\NAT@citex}
  {\@citea\NAT@hyper@{%
     \NAT@nmfmt{\NAT@nm}%
     \hyper@natlinkbreak{\NAT@aysep\NAT@spacechar}{\@citeb\@extra@b@citeb}%
     \NAT@date}}
  {\@citea\NAT@nmfmt{\NAT@nm}%
   \NAT@aysep\NAT@spacechar\NAT@hyper@{\NAT@date}}{}{}

\patchcmd{\NAT@citex}
  {\@citea\NAT@hyper@{%
     \NAT@nmfmt{\NAT@nm}%
     \hyper@natlinkbreak{\NAT@spacechar\NAT@@open\if*#1*\else#1\NAT@spacechar\fi}%
       {\@citeb\@extra@b@citeb}%
     \NAT@date}}
  {\@citea\NAT@nmfmt{\NAT@nm}%
   \NAT@spacechar\NAT@@open\if*#1*\else#1\NAT@spacechar\fi\NAT@hyper@{\NAT@date}}
  {}{}
  
\usepackage{array}
\newcolumntype{C}[1]{>{\centering\let\newline\\\arraybackslash\hspace{0pt}}m{#1}}

\def\aj{AJ}
\def\araa{ARA\&A}
\def\apj{ApJ}
\def\apjl{ApJ}
\def\apjs{ApJS}
\def\ao{Appl.~Opt.}
\def\apss{Ap\&SS}
\def\aap{A\&A}

\def\mnras{MNRAS}

\def\pasp{PASP}
\def\pasj{PASJ}

\def\nat{Nature}

\def\procspie{Proc.~SPIE}

\def\arcdeg{\hbox{$^\circ$}}

\def\pg{PG\,1211+143\xspace}

\newcommand{\arcus}{\textit{Arcus}\xspace}
\newcommand{\xmm}{\textit{XMM-Newton}\xspace}
\newcommand{\chandra}{\textit{Chandra}\xspace}
\newcommand{\hetgs}{{HETGS}\xspace}
\newcommand{\heg}{{HEG}\xspace}
\newcommand{\meg}{{MEG}\xspace}
\newcommand{\herschel}{\textit{Herschel Space Observatory}\xspace}
\newcommand{\hubble}{\textit{Hubble Space Telescope}\xspace}
\newcommand{\hst}{\textit{HST}\xspace}
\newcommand{\hcos}{{COS}\xspace}
\newcommand{\hfos}{{FOS}\xspace}
\newcommand{\stis}{{STIS}\xspace}
\newcommand{\hstcos}{\textit{HST}-COS\xspace}
\newcommand{\hstfos}{\textit{HST}-FOS\xspace}

\newcommand{\jvla}{{VLA}\xspace}

\newcommand\ciao{\textsc{ciao}\xspace}
\newcommand\isis{\textsc{isis}\xspace}
\newcommand\xspec{\textsc{xspec}\xspace}
\newcommand\xstar{\textsc{xstar}\xspace}
\newcommand\mpixstar{\textsc{mpi\_xstar}\xspace}
\newcommand\odyssey{\textsc{odyssey}\xspace}

\newcommand\caldb{\textsc{caldb}\xspace}

\newcommand\mkgrmf{\textsf{mkgrmf}\xspace}

\newcommand\fullgarf{\textsf{fullgarf}\xspace}
\newcommand\tgextract{\textsf{tgextact}\xspace}
\newcommand\combine{\textsf{combine\_datasets}\xspace}
\newcommand\aglc{\textsf{aglc}\xspace}
\newcommand\group{\textsf{group}\xspace}

\newcommand\tbnew{\textsf{tbnew}\xspace}
\newcommand\highecut{\textsf{highecut}\xspace}
\newcommand\diskbb{\textsf{diskbb}\xspace}
\newcommand\zpowerlw{\textsf{zpowerlw}\xspace}
\newcommand\zgauss{\textsf{zgauss}\xspace}
\newcommand\gabs{\textsf{gabs}\xspace}
\newcommand\warmabs{\textsf{warmabs}\xspace}
\newcommand\xstarabs{\textsf{xstar\_absorber}\xspace}
\newcommand\emcee{\textsf{emcee hammer}\xspace}

\newcommand{\sii}{S\,{\sc ii}}

\newcommand{\nv}{N\,{\sc v}}

\newcommand{\hi}{H\,{\sc i}}

\newcommand{\nex}{Ne\,{\sc x}}
\newcommand{\neix}{Ne\,{\sc ix}}
\newcommand{\mgxii}{Mg\,{\sc xii}}
\newcommand{\mgxi}{Mg\,{\sc xi}}
\newcommand{\sixiv}{Si\,{\sc xiv}}
\newcommand{\sixiii}{Si\,{\sc xiii}}

\newcommand{\fexxvi}{Fe\,{\sc xxvi}}
\newcommand{\fexxv}{Fe\,{\sc xxv}}

\newcommand{\lya}{Ly$\alpha$}

\newcommand{\hea}{He$\alpha$}

\newcommand{\bc}{\color{blue}}

\makeatletter
\patchcmd{\frontmatter@RRAP@format}{(}{}{}{}
\patchcmd{\frontmatter@RRAP@format}{)}{}{}{}
\renewcommand\Dated@name{}
\makeatother

\newcommand{\altaffilmarkc}[1]{\altaffilmark{\bc #1}}
\newcommand{\emailc}[1]{{\bc #1}}

\shorttitle{Ultra-fast Outflow of \pg}
\shortauthors{Danehkar et al.}

\begin{document}

\title{The Ultra-Fast Outflow of the Quasar PG\,1211+143 \\ as Viewed by Time-Averaged \chandra Grating Spectroscopy}

\author{Ashkbiz~Danehkar\altaffilmarkc{1}, 
Michael~A.~Nowak\altaffilmarkc{2}, 
Julia~C.~Lee\altaffilmarkc{1,3}, 
Gerard~A.~Kriss\altaffilmarkc{4}, 
Andrew~J.~Young\altaffilmarkc{5}, 
Martin~J.~Hardcastle\altaffilmarkc{6}, 
Susmita~Chakravorty\altaffilmarkc{7}, 
Taotao~Fang\altaffilmarkc{8}, 
Joseph~Neilsen\altaffilmarkc{9}, 
Farid~Rahoui\altaffilmarkc{10}, and  
Randall~K.~Smith\altaffilmarkc{1}
}

\affil{\altaffilmark{1}\,Harvard-Smithsonian Center for Astrophysics,
  60 Garden Street, Cambridge, MA 02138, USA;
  \emailc{ashkbiz.danehkar@cfa.harvard.edu}\\
%\hyperref[mailto:ashkbiz.danehkar@cfa.harvard.edu]{ashkbiz.danehkar@cfa.harvard.edu}\\
\altaffilmark{2}\,Massachusetts Institute of Technology, Kavli
Institute for Astrophysics, Cambridge, MA 02139, USA \\
\altaffilmark{3}\,Harvard John A. Paulson School of Engineering and 
Applied Science, 29 Oxford Street, Cambridge, MA 02138 USA \\
\altaffilmark{4}\,Space Telescope Science Institute, 3700 San Martin
Drive, Baltimore, MD, 21218, USA \\
\altaffilmark{5}\,University of Bristol, H. H. Wills Physics
Laboratory, Tyndall Avenue, Bristol BS8 1TL, UK\\
\altaffilmark{6}\,University of Hertfordshire, School of Physics,
Astronomy and Mathematics, Hatfield, Hertfordshire AL10 9AB, UK\\
\altaffilmark{7}\,Department of Physics, Indian Institute of Science, Bangalore 560012, India \\
\altaffilmark{8}\,Xiamen University, Institute for Theoretical Physics
and Astrophysics, Department of Astronomy, Xiamen, Fujian 361005,
China\\
\altaffilmark{9}\,Villanova University, Mendel Hall, Room 263A,
800 E. Lancaster Avenue, Villanova, PA 19085, USA  \\
\altaffilmark{10}\,European Southern Observatory, Karl
Schwarzschild-Stra{\ss}e 2, D-85748 Garching bei M\"{u}nchen, Germany
}

\date[ ]{\footnotesize\textit{Received 2017 October 17; revised 2017 December 14; accepted 2017 December 18; published 2018 February 2}}

\begin{abstract}
We present a detailed X-ray spectral study of the quasar PG\,1211+143 based on \textit{Chandra} High Energy Transmission Grating Spectrometer (HETGS) observations collected in a multi-wavelength campaign with UV data using the \textit{Hubble Space Telescope} Cosmic Origins Spectrograph (\textit{HST}-COS) and radio bands using the Jansky Very Large Array (VLA). We constructed a multi-wavelength ionizing spectral energy distribution using these observations and archival infrared data to create {\sc xstar} photoionization models specific to the PG\,1211+143 flux behavior during the epoch of our observations.  Our analysis of the \textit{Chandra}-HETGS spectra yields complex absorption lines from H-like and He-like ions of Ne, Mg and Si which confirm the presence of an ultra-fast outflow (UFO) with a velocity $\sim-$17\,300~km\,s$^{-1}$ (outflow redshift $z_{\rm out} \sim -0.0561$) in the rest frame of PG\,1211+143.   This absorber is well described by an ionization parameter $\log \xi \sim2.9$~erg\,s$^{-1}$\,cm and column density $\log N_{\rm H} \sim21.5$~cm$^{-2}$. This corresponds to a stable region of the absorber's thermal stability curve, and furthermore its implied neutral hydrogen column is broadly consistent with a broad Ly$\alpha$ absorption line at a mean outflow velocity of $\sim -16\,980$~km\,s$^{-1}$ detected by our \textit{HST}-COS observations. Our findings represent the first simultaneous detection of a UFO in both X-ray and UV observations.  Our VLA observations provide evidence for an active jet in PG\,1211+143, which may be connected to the X-ray and UV outflows; this possibility can be evaluated using very-long-baseline interferometric observations.
\end{abstract}

\keywords{ quasars: absorption lines --- quasars: individual
  (\pg) --- galaxies: active --- galaxies: Seyfert ---
  X-rays: galaxies}

\section{Introduction}
\label{pg1211:introduction}

X-ray observations of active galactic nuclei (AGNs) reveal blueshifted
absorption features, which have been interpreted as outflows of
photoionized gas along the line of sight \citep{Halpern1984}. Soft
X-ray absorption lines are commonly referred to as warm absorbers
(WAs), while those ionized absorbers with a velocity higher than
10\,000 km\,s$^{-1}$ are defined as ultra-fast outflows
\citep[UFOs;][]{Tombesi2010a}. WAs have been observed in over half
of Seyfert 1 galaxies
\citep[e.g.,][]{Reynolds1995,Reynolds1997,George1998,Laha2014}, which exhibit
outflow velocities in the range of 100--500 km\,s$^{-1}$
\citep[e.g.,][]{Kaspi2000,Blustin2002,McKernan2007}. On the other
hand, X-ray observations of iron absorption lines can indicate outflow
velocities that are quite large, up to mildly relativistic values of
$\sim 0.1$--$0.4c$
\citep[e.g.,][]{Pounds2003,Cappi2006,Braito2007,Cappi2009}. More
recent studies show that UFOs are identified in a significant fraction
($\sim 30$ per cent) of radio-quiet and radio-loud AGNs
\citep{Tombesi2010a,Tombesi2011,Tombesi2012,Tombesi2014}. Recently,
\citet{Tombesi2013} concluded that UFOs and WAs are associated with
different locations of a single large-scale stratified outflow in the
AGN, suggesting a unified model for accretion powered sources
\citep{Kazanas2012}. However, \citet{Laha2014,Laha2016} instead suggested that
UFOs and WAs may be associated with two different outflows with
distinctive physical conditions and outflow velocities.

\begin{table*}
%\centering
\begin{center}
\caption[]{Observation log of \pg}
\label{tab:obs:log}
\begin{tabular}{lllcccccc}
  \hline\hline\noalign{\smallskip}
Observatory & Detector  & Gratings  & Seq./PID & Obs.ID &  UT Start  &  UT End 
            & Time (ks) \\
%\noalign{\smallskip}
%(1) & (2) &  (3) & (4) & (5) & (6) & (7) & (8)  \\
\noalign{\smallskip}
   \hline\noalign{\smallskip}
\chandra & ACIS-S & \hetgs & 703109 & 17109 & 2015 Apr 09, 08:22 & 2015 Apr 10, 14:32 
         & 104.68 \\
\noalign{\smallskip}
\chandra & ACIS-S & \hetgs & 703109 & 17645 & 2015 Apr 10, 17:55 & 2015 Apr 11, 06:56 
         & 44.33 \\
\noalign{\smallskip}
\chandra & ACIS-S & \hetgs & 703109 & 17646 & 2015 Apr 12, 02:04 & 2015 Apr 13, 02:15 
         & 83.65 \\
\noalign{\smallskip}
\chandra & ACIS-S & \hetgs & 703109 & 17647 & 2015 Apr 13, 13:54 & 2015 Apr 14, 02:09 
          & 42.22 \\
\noalign{\smallskip}
\chandra & ACIS-S & \hetgs & 703109 & 17108 & 2015 Apr 15, 07:13 & 2015 Apr 16, 03:10 
         & 68.89 \\
\noalign{\smallskip}
\chandra & ACIS-S & \hetgs & 703109 & 17110 & 2015 Apr 17, 06:40 & 2015 Apr 18, 08:28 
         & 89.56 \\
\noalign{\medskip}
\hst     & \hcos    & G140L & 13947 & LCS501010  & 2015 Apr 12, 15:50 & 2015 Apr 12, 16:28 
         & 1.90 \\
\noalign{\smallskip}
\hst    & \hcos    & G140L & 13947 & LCS504010  & 2015 Apr 14, 13:52 & 2015 Apr 14, 14:30 
        & 1.90 \\
\noalign{\smallskip}
\hst     & \hcos    & G140L & 13947 & LCS502010  & 2015 Apr 14, 15:37 & 2015 Apr 14, 16:15 
         & 1.90 \\
\noalign{\smallskip}
\hst    & \hcos    & G130M & 13947 & LCS502020  & 2015 Apr 14, 17:17 & 2015 Apr 14, 19:05 
        & 2.32 \\
\noalign{\medskip}
\hst    & \hfos    & G130H & 1026 & Y0IZ0304T  & 1991 Apr 13, 08:21 & 1991 Apr 13, 08:56 
        & 2.00 \\
\noalign{\smallskip}
\hst     & \hfos    & G130H & 1026 & Y0IZ0305T  & 1991 Apr 16, 09:56 & 1991 Apr 16, 10:31 
         & 2.00 \\
\noalign{\smallskip}
\hst     & \hfos    & G270H & 1026 & Y0IZ0404T  & 1991 Apr 16, 07:51 & 1991 Apr 16, 07:57 
         & 3.49 \\
\noalign{\smallskip}
\hst     & \hfos    & G190H & 1026 & Y0IZ0406T  & 1991 Apr 16, 09:00 & 1991 Apr 16, 09:25 
         & 1.34 \\
\noalign{\smallskip}\hline
%\noalign{\smallskip}
\end{tabular}
\end{center}
\begin{tablenotes}
\item[1]\textbf{Notes.} The table above lists information for each of
  the three observations of \pg used in this work, namely \chandra-\hetgs
  (Seq. 703109), \hstcos (PID 13947), and \hstfos (PID
  1026). The columns list the observatory name, spectrometer
  instrument, grating setting, program ID or sequence, observation ID,
  start time, end time, and total exposure duration.
\end{tablenotes}
\end{table*}

The optically bright quasar \pg in a nearby, luminous narrow line
Seyfert 1 galaxy \citep[$z = 0.0809$;][]{Marziani1996,Rines2003} is
one of the AGNs with potentially mildly relativistic UFOs
\citep{Pounds2003,Pounds2006,Fukumura2015,Pounds2016a}. Over a decade
ago, \citet{Pounds2003} reported absorption lines of H- and He-like
ions of C, N, O, Ne, Mg, S and Fe with an outflow velocity of
$\sim -24$,000 km\,s$^{-1}$ ($\sim  -0.08c$).\footnote{
For prior work on \pg, we are assuming that all references to
velocities are true velocities in the rest frame of \pg, with
$v$ in $\rm km~s^{-1}$ converted to $zc$ simply by dividing by
$c=2.9979 \times 10^5~\rm km~s^{-1}$.}
Moreover, \citet{Reeves2005} reported the detection of {\it redshifted} H-like
or He-like iron absorption lines with velocities in the range of
$0.2c$--$0.4c$, which could be evidence for pure gravitational
redshift by the supermassive black hole (SMBH).  The presence of UFOs
in \pg was challenged by \citet{Kaspi2006}; however, they were again
confirmed by later works
\citep{Pounds2006,Pounds2007,Pounds2009,Tombesi2010a,Tombesi2011}. More
recently, a second high-velocity component with $\sim -0.066c$ ($-19\,800~\rm km\,s^{-1}$) 
was detected, in addition to a confirmation of a previously identified
higher velocity component of $\sim -0.129c$ ($-38\,700~\rm km\,s^{-1}$) 
\citep{Pounds2014,Pounds2016a,Pounds2016b}. \hubble (\hst) UV
observations of \pg taken with the Space Telescope Imaging
Spectrograph (\stis) had also revealed the presence of four strong absorbers
at observed redshifts of $0.01649$ to $0.02586$ 
\citep[implied outflow velocities of $-$15\,650 to $-$18\,400 $\rm km~s^{-1}$;][]{Penton2004,Tumlinson2005, Danforth2008,Tilton2012}. 
These authors postulated that these could be attributed to the intergalactic
medium (IGM) or outflows from unseen satellite galaxies 
(see \S\,\ref{pg1211:results:uv}).

Many of these disparate results can be explained by the apparent highly variable
nature of the UFO phenomenon. Long, intensive observations of AGN such as
IRAS\,13224$-$3809 \citep{Parker2017a,Parker2017b} and PDS\,456
\citep{Matzeu2016} show UFO variability on timescales of 10,000 to 100,000 s.
The character of the absorption also depends on the state of the illuminating
X-ray source, with the outflowing gas often showing an ionization response
\citep{Parker2017a}, or a correlation between ionization state, outflow
velocity, and X-ray flux \citep{Matzeu2017,Pinto2017}.
While these characteristics suggest radiative acceleration of the outflow
\citep{Matzeu2017}, other authors offer a more complex vision of these
observationally complex winds.
\cite{Konigl1994} described a magnetohydrodynamical wind model as a
possible explanation for the warm absorber winds observed in many AGN.
\cite{Fukumura2010} and \cite{Kazanas2012} adapted magnetohydrodynamical winds
to high-velocity winds launched from the accretion disk, compatible with UFOs.
These winds could have a complex structure, with velocity and ionization state
dependent upon the observer's line of sight. In addition, once launched,
such disk winds would be photoionized by the central X-ray source and subject to
additional acceleration due to radiation pressure.

While \cite{Fukumura2010b} did attempt a theoretical demonstration that
the X-ray absorption by Fe\,{\sc xxv} can co-exist with broad
ultraviolet absorption by C\,{\sc iv}, their scenario required a very
low X-ray to UV luminosity ratio ($\alpha_{ox} \sim 2$) in order to
keep the UV ionization of the gas low. PG\,1211+143 has a fairly high
X-ray to UV luminosity ratio, however, with $\alpha_{ox} = 1.47$, so
although C\,{\sc iv} absorption might not be expected, trace amounts
of H\,{\sc i} can remain, even in very highly ionized gas.
Our observation is the first to detect an X-ray UFO observed
simultaneously with UV broad absorption in H\,{\sc i} Ly$\alpha$.

As an alternative to an outflowing wind producing the UFO features
in PG\,1211+143, \cite{Gallo2013} presented a model in which the broad
absorption is produced by blurred reflection in the X-ray illuminated
atmosphere of the accretion disk. Again, in such a model, trace amounts
of H\,{\sc i} may still be present, which could give rise to similar
UV absorption features. 

Our main goal in this work is to identify and characterize 
 the blueshifted X-ray absorption features of \pg based on our \chandra\ 
observations, which are part of a program that includes simultaneous
\hst UV and Jansky Very Large Array (VLA) radio observations.
Given the relativistic velocities of the outflows we are examining, it
is important to use a full relativistic treatment for all velocities and
redshifts. For clarity in understanding the nomenclature we use in this paper,
we summarize the following definitions for quantities we will use: $z_{\rm rest}$ is the rest frame redshift of the host galaxy ($z_{\rm rest}=0.0809$ for \pg), $z_{\rm obs}$ is the observed redshift (in our reference frame)
of a spectral feature, $z_{\rm out}$ is the redshift of an outflow in the frame of \pg, $v_{\rm out}$ is the velocity of an outflow in the frame of \pg, $\lambda_{\rm obs}$ is the observed wavelength of a spectral feature, $\lambda_0$ is the rest wavelength (vacuum) of a spectral feature. These quantities are related by the usual special relativistic formulations: $z_{\rm obs} = (\lambda_{\rm obs}/\lambda_0) - 1$, $z_{\rm out} = (1+z_{\rm obs})/(1+z_{\rm rest}) - 1$,
$v_{\rm out} = c [(1+z_{\rm out})^2-1]/[(1+z_{\rm out})^2+1]$, and $z_{\rm out} = \sqrt{[(1+v_{\rm out}/c)/(1-v_{\rm out}/c)]} - 1$, where $c$ is the speed of light.

This paper is focused primarily on the X-ray analysis, and organized
as follows. Section~\ref{pg1211:observations} describes the
observations and data reduction. In \S\,\ref{pg1211:lcs} we inspect the
X-ray light curve and hardness ratios in order to see
whether all spectra can be co-added for analysis. 
In \S\,\ref{pg1211:analysis} we model
the X-ray continuum and Fe emission lines. In
\S\,\ref{pg1211:modeling} we describe in detail the modeling of the
ionized absorber using the photoionization code \xstar. 
The \hst UV results are reported in a complementary paper \citep{Kriss2017}.
A summary of its findings as relevant to this paper are 
presented in \S\ref{pg1211:results:uv}. 
Section~\ref{pg1211:jet} presents the results of our VLA observations.
In \S\,\ref{pg1211:discussion}, we discuss the implications of 
our X-ray and UV absorption features. 
Finally, we summarize our results in \S\,\ref{pg1211:conclusions}.

\section{Observations}
\label{pg1211:observations}

\subsection{\chandra-\hetgs Observation}
\label{pg1211:observations:hetg}

We observed \pg over six visits (Proposal 16700515, PI: Lee) from 2015
April 9 (MJD 57121.362) to April 17 (MJD 57130.351) with the High
Energy Transmission Grating Spectrometer
\citep[\hetgs;][]{Weisskopf2002,Canizares2005} using the \chandra\ 
Advanced CCD Imaging Spectrometer \citep[ACIS;][]{Garmire2003}. The
observation log is listed in Table \ref{tab:obs:log}. The total useful
exposure time of the observations was about $433$\,ks, with individual
exposure times ranging from 42--105\,ks. 
The \hetgs has two grating assemblies, the medium energy grating
(\meg) and the high energy grating (\heg). The \meg has 
a full-width at half-maximum (FWHM) resolution of 0.023 {\AA} and covers 0.4--7\,keV, 
whereas the \heg has a FWHM resolution of 0.012 {\AA} and covers
0.8--10\,keV. 
The
\meg response is therefore more efficient in the soft band, while the
\heg can more effectively measure the hard band.

We used the \ciao package \citep[v.\,4.8;][]{Fruscione2006},
along with calibration files from the \caldb (v 4.7.0), to
process the \hetgs data in the standard way, which produces the plus and
minus first-order ($m=\pm 1$) \meg and \heg data and the response
files. Spectra 
files
were extracted using the \ciao tool \tgextract from
the $-1$ and $+1$ arms of the \meg and \heg. 
Redistribution and response files were generated using the  \ciao tools \mkgrmf and \fullgarf, respectively.

We regridded the \heg spectra to match the \meg bins, and then combined
the \heg and \meg data using the \combine function in
the Interactive Spectral Interpretation System (\isis)
package\footnote{\url{http://space.mit.edu/asc/isis/}} v.\,1.6.2-35
\citep{Houck2000} for spectral fitting.  We further rebinned the
combined data (see discussion in the Appendix), starting at 0.4 keV, to a
minimum signal-to-noise of 4 and a minimum of 4 spectral
channels per bin (i.e., approximately the spectral resolution of the
\meg detector).  The signal-to-noise criterion determined the binning
below $\sim 1$\,keV, while the minimum channel criterion determined
the binning above $\sim 1$\,keV.  While not ideal, this was necessary
for the signal-to-noise required of our analysis, although
a blind-line search with uniform \meg binning
is consistent with the results presented in this paper (see Appendix). We fit our
spectral models only in the 0.5--6.75 keV range (all energy ranges
above refer to energies in the observer frame), which corresponds to
the 0.54--7.3 keV range in the rest frame.

\subsection{UV Observations}
\label{pg1211:observations:cos}

Simultaneous \hst far-ultraviolet (FUV) spectra of \pg listed in Table
\ref{tab:obs:log} were obtained on 2015 April 12 and 14 over four
observations (PID 13947, PI: Lee) using the Cosmic Origins
Spectrograph \citep[\hcos;][]{Osterman2011,Green2012}.
We used grating G140L to cover the 912--2000\,{\AA}
wavelength range.  To fill in the gap from 1190--1270 \AA\ between the two
segments of the \hcos detector, we used grating G130M during the second visit.
See \S\ref{pg1211:results:uv} for a summary of the observational results,
and \cite{Kriss2017} for details of the data reduction, calibration
and analysis.

The earlier \hst UV observations of \pg listed in Table
\ref{tab:obs:log} were taken on 1991 April 13 and 16 (PID 1026, PI:
Burbidge) using the Faint Object Spectrograph
\citep[\hfos;][]{Keyes1995}. The \hfos data were retrieved from the
\hst legacy archive, which include two exposures of 2.0\,ks
made in the G130H (1150--1600\,{\AA}), one exposure of 1.34\,ks in the
G190H (1580--2330\,{\AA}), and one exposure of 3.49\,ks in the G270H
(2220--3300\,{\AA}).   

\subsection{Radio Observations}
\label{pg1211:observations:radio}

Radio data of PG 1211+143 at C-band (4-8 GHz) and K-band
(18-26 GHz) were taken on 2015 June 18 with the `A' configuration
of the Karl G. Jansky Very Large Array (VLA) 
for a total time on
source of 7.0 ks at K-band and 3.9 ks at C-band. At K-band we made
frequent visits to a bright nearby calibrator, J1215+1654, to ensure
good phase calibration, and also used pointing reference mode. The
data were reduced in the standard manner for continuum mode using the 
Common Astronomy Software Applications (CASA) package
to produce broad-band images with nominal central frequencies of 6 and
22 GHz, with resolution of respectively $0\farcs24 \times 0\farcs22$
and $0\farcs065 \times 0\farcs61$, and rms noise levels of
respectively 10 and 6 $\mu$Jy beam$^{-1}$: in addition we split the
K-band bandwidth into low and high frequencies to make 20-GHz and
24-GHz images. We also analysed a short archival observation with the
un-upgraded \jvla, taken on 1993 May 02 in the `B' configuration as part
of program AP250, which was reduced in the standard manner using
the Astronomical Image Processing System (AIPS).

\section{X-ray Brightness Changes} 
\label{pg1211:lcs}

\begin{figure}
\begin{center}
\includegraphics[width=3.2in, trim = 50 20 0 0, clip, angle=0]{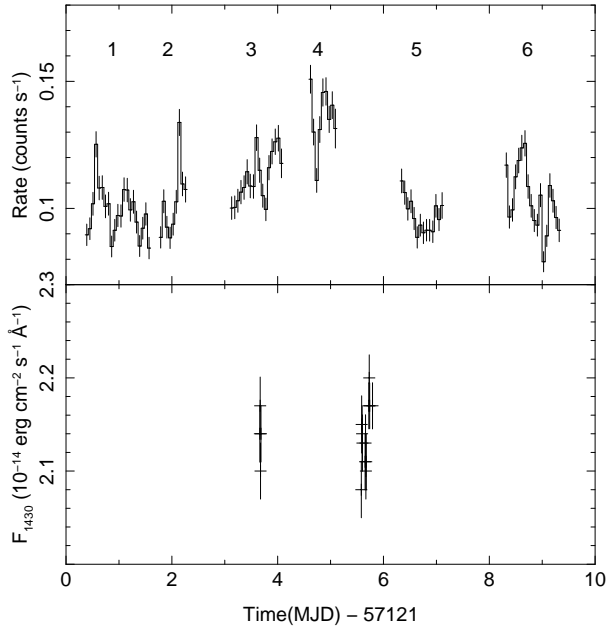}%
\caption{Upper panel: long-term light curve of \pg in the 0.4--8 keV
  broad band observed with the \chandra-\hetgs from 2015 April 9 to 17
  (MJD: 57121.362--57130.351) binned using 5000-s time
  intervals. Lower panels: \hstcos sampling light curve.  The time
  unit is day and zero corresponds to MJD 57121.
\label{fig:pg1211:light}%
}
\end{center}
\end{figure}

Figure \ref{fig:pg1211:light} (upper panel) displays the 0.4--8 keV
light curves of \pg measured by the \hetgs, which were obtained
from a combination of the \meg and \heg data. The observation numbers
are labeled (with ``1--6'') according to the time sequence of the
exposures listed in the first column of Table \ref{tab:obs:log}. The
source appeared to be at its lowest brightness at the beginning, while
it reached the highest brightness on April 13 (4th observation). 
It then returned to a lower brightness on April 15. As can be seen, there
were only moderate changes in brightness over the 9-day observation.

Figure \ref{fig:pg1211:light} (lower panel) displays the FUV continua
of \pg measured by the \hstcos. The G140L continuum light curve has an
average flux of $F_{\lambda}$(1430\,{\AA}$)=(2.14 \pm 0.09) \times
10^{-14}$ erg\,cm$^{-2}$\,s$^{-1}$\,{\AA}$^{-1}$, and shows very small
variations in two days, which is consistent with the G130M continuum
$F_{\lambda}$(1430\,{\AA}$)=2.18 \times 10^{-14}$
erg\,cm$^{-2}$\,s$^{-1}$\,{\AA}$^{-1}$. The G140L data combined with
X-ray, infrared, and radio were used for constructing the ionizing SED
for the photoionization modeling discussed in
\S\,\ref{pg1211:results:sed}.

\subsection{X-ray Spectral Hardness}
\label{pg1211:hardness}

We begin with hardness ratio analysis to determine whether we can
reasonably co-add all spectra for analysis.  We used the \aglc
program\footnote{\url{http://space.mit.edu/cxc/analysis/aglc/}}
originally developed for the \chandra\ Transmission Grating
Data Catalog \citep[TGCat;][]{Huenemoerder2011} to create light curves
in 5000-s bins for three bands: the soft band (\textit{S}: 0.4--1.1
keV), the medium band (\textit{M}: 1.1--2.6 keV), and the hard band
(\textit{H}: 2.6--8 keV). The \meg data were used to compute the light
curves in the \textit{S} band, while the light curves in the
\textit{M} and \textit{H} bands were extracted from a combination of
the \meg and \heg data over the spectral regions of 0.4--4 and 4--8 keV,
respectively. Positive and negative first-order spectra for \heg and
\meg were also combined to generate the light curves of the broad band
on the 0.4--8 keV spectral region (the light curve in
Figure\,\ref{fig:pg1211:light}, upper panel).

\begin{figure}
\begin{center}
\includegraphics[width=3.0in, trim = 50 20 0 0, clip, angle=0]{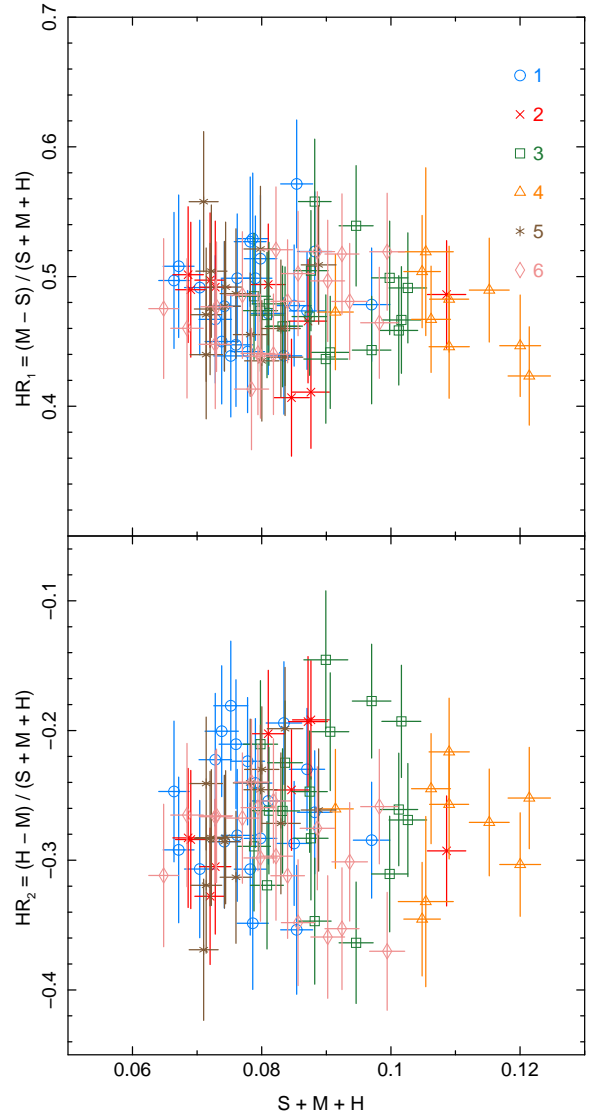}%
\caption{The hardness ratios $\mathrm{HR_{1}}=(M-S)/(S+M+H)$ (upper
  panel), and $\mathrm{HR_{2}}=(H-M)/(S+M+H)$ (lower panel) plotted
  against the light curve of all the energy bands ($S+M+H$).
\label{fig:pg1211:hard}%
}
\end{center}
\end{figure}

We define the count rate hardness ratios ($\mathrm{HR_{1}}$ and
$\mathrm{HR_{2}}$) using the soft (\textit{S}), medium (\textit{M}),
and hard (\textit{H}) light-curve bands as follows:
\begin{align}
\mathrm{HR_{1}} &= \frac{M-S}{S+M+H},\label{eq_1}\\
\mathrm{HR_{2}} &= \frac{H-M}{S+M+H},\label{eq_2}
\end{align}
The uncertainties on the hardness ratios are estimated by propagating
the band errors assuming Gaussian statistics.  By these definitions,
as the X-ray spectra become harder, the hardness ratio
$\mathrm{HR_{2}}$ increases.  The disk blackbody is expected to
dominate the soft band,  
while the powerlaw and warm
absorber are the dominant features in the medium band. 
We note that the powerlaw dominates the hard band. 

The hardness ratios $\mathrm{HR_{1}}$ and $\mathrm{HR_{2}}$ are
plotted against the sum of the light curve of all the energy bands
($S+M+H$) in Figure \ref{fig:pg1211:hard}. The hardness ratio
$\mathrm{HR_{1}}$ increases with the count rate of the \textit{M}
band, and decreases with the count rate of the \textit{S} band,
so a decrease in $\mathrm{HR_{1}}$ may be related to a stronger soft
excess of an accretion disk. 
In contrast, the hardness ratio $\mathrm{HR_{2}}$ decreases with the
count rate of the \textit{M} band and increases with the count rate of
the \textit{H} band, so an increase in $\mathrm{HR_{2}}$ may indicate
a hardening of the powerlaw spectral index.  The source tends to be
slightly softer during the 4th observation (on April 13) when it has
the highest luminosity. The variability of the source will be studied
in a separate paper. For the time-averaged analysis, we excluded the
4th observation from the combined observations, and restrict ourselves
to the five lowest flux spectra. The time-averaged spectrum of \pg is
obtained from combining $\sim 390$\,ks of exposure time.

\section{X-ray Spectral Analysis}
\label{pg1211:analysis}

We performed our X-ray spectral analysis in several steps to ensure
consistent and optimal results. We assessed in \S\,\ref{pg1211:hardness} 
whether the spectra can be reasonably co-added. To model the continuum, we used a phenomenological model
consisting of an accretion disk model (soft band), a power-law model
(hard band), and photoelectric absorption models
(\S\,\ref{pg1211:results:continuum}). For the absorption and emissions
lines, we measure their properties first with Gaussian emission
(\S\,\ref{pg1211:iron:emissions}) and absorption functions
(\S\,\ref{pg1211:iron:absorptions}) superimposed on the continuum.
The analysis conducted in this section is intended as an additional measure
to ensure that our reported results are robust to both simple (this section),
and complex (\S\ref{pg1211:modeling}) analysis techniques.

\begin{table}
\caption{Best-fit Parameters for the Continuum Model and Fe Lines.
\label{pg1211:fit:continuum}
}
\centering
\begin{tabular}{lC{3.5cm}c}
\hline\hline
\noalign{\smallskip}
{Component} & {Parameter} & {Value}    \\
\noalign{\smallskip}
\hline 
\noalign{\smallskip}
 & $\chi^{2}/{\rm d.o.f}$ \dotfill &  $ 739/659$ \\ 
\noalign{\medskip}  
\tbnew & $N_{\rm H}$(cm$^{-2}$ ) \dotfill & $     2.32^{+0.48}_{-0.55} \times 10^{21}$ \\ 
\noalign{\medskip} 
\highecut & $E_{c}$(keV) \dotfill  & $    4.55^{+1.16}_{-2.01}$   \\ 
\noalign{\smallskip}
                  & $E_{f}$(keV) \dotfill & $   13.47^{}_{-10.85}$    \\ 
\noalign{\medskip}                   
\diskbb & $T_{\rm in}$(keV) \dotfill & $     0.064^{+0.004}_{-0.003}$  \\ 
\noalign{\smallskip}
                  & $R_{\rm in}$(pc) \dotfill  & $    15.96^{+16.80}_{-8.25} \times 10^{-4}$   \\ 
\noalign{\medskip} 
\zpowerlw & $\Gamma$ \dotfill  & $     1.77^{+0.04}_{-0.19}$   \\ 
\noalign{\smallskip}
                  & $F$(erg\,cm$^{-2}$\,s$^{-1}$) \dotfill & $    5.45^{+0.12}_{-0.07} \times 10^{-12}$    \\ 
\noalign{\medskip} 
\textsf{zgauss$_{\rm K\alpha}$} & $E$(keV) \dotfill & $     6.41$   \\ 
\noalign{\smallskip}
(Fe\,K$\alpha$)                 & $\sigma$(keV) \dotfill & $     0.04$   \\ 
\noalign{\smallskip}
                & $F$(erg\,cm$^{-2}$\,s$^{-1}$) \dotfill  & $     3.35^{+2.31}_{-2.03}\times10^{-14}$ \\ 
\noalign{\medskip}       
\textsf{zgauss$_{\rm He\alpha}$} & $E$(keV) \dotfill & $     6.67$   \\ 
\noalign{\smallskip}
(\fexxv)                          & $\sigma$(keV) \dotfill & $     0.06$   \\ 
\noalign{\smallskip}
                & $F$(erg\,cm$^{-2}$\,s$^{-1}$) \dotfill  & $     4.87^{+3.27}_{-2.74}\times10^{-14}$ \\ 
\noalign{\medskip}  
\textsf{zgauss$_{\rm Ly\alpha}$} & $E$(keV) \dotfill & $     6.96$   \\ 
\noalign{\smallskip}
(\fexxvi)                         & $\sigma$(keV) \dotfill & $     0.06$   \\ 
\noalign{\smallskip}
                & $F$(erg\,cm$^{-2}$\,s$^{-1}$) \dotfill  & $     1.67^{+2.97}_{}\times10^{-14}$ \\ 
\noalign{\smallskip}
\hline
\noalign{\smallskip}
\end{tabular}
\begin{tablenotes}
\item[1]\textbf{Notes.} Redshift is fixed to $z=0.0809$.

The Fe K$\beta$ line at 7.05\,keV was estimated by tying its relative
energy and absolute width to the K$\alpha$ line, and its flux to a
K$\beta$/K$\alpha$ flux ratio of 0.135. The apparent inner disk radius
($R_{\rm in}$) is calculated from the norm value \citep[see disk
  equations in][]{Makishima1986,Kubota1998}, adopting the luminosity
distance of $D=358$\,Mpc and the inclination angle of
$\theta=28^{\circ}$ \citep{Zoghbi2015}. All errors are quoted at the
90\% confidence level. The upper error of the \highecut folding energy
($E_f$) is at the imposed $300$\,keV limit. The lower error of the
\textsf{zgauss$_{\rm Ly\alpha}$} flux ($F$) is at $0$. The use
of a \tbnew component here is to enable greater flexibility in the
phenomenological description of the continuum shape, and should not be
attributed to that amount of cold absorption in our Galaxy nor in that of
\pg.
\end{tablenotes}
\end{table}

\begin{table*}
\begin{center}
  \caption[]{X-ray Absorption Lines in \pg identified with Gaussian
    absorption functions (\gabs)}
   \label{tab:list:absorption}
   \begin{tabular}{lcccccccc}
  \hline\hline\noalign{\smallskip}
Line  & $E_{\rm lab}$ & $\lambda_{\rm lab}$ & $E_{\rm rest}$  & $\lambda_{\rm rest}$  
      & $v_{\rm out}$  & $W_{\lambda}$ & $v_{\rm turb}$         &  $\log N_{j}$  \\
      &   (keV)      &  ({\AA})           &  (keV)         &  ({\AA})     
      & (km s$^{-1}$)          & (m\AA)   & (km s$^{-1}$) &  (cm$^{-2}$)   \\
\noalign{\smallskip}
\hline\noalign{\smallskip}
\nex\ \lya	& 1.022	 & 12.132  & 1.081  & 11.469 &$-16750^{+180}_{-630}$	
                &$41.0 \pm 6.9$	&$280^{+460}_{-190}$	&$17.4^{+0.1}_{-0.1}$\\
\noalign{\smallskip}
\neix\ \hea	&0.923	& 13.433 &  0.976 & 12.703 &$-16660^{+1290}_{-360}$	
                &$7.5\pm 6.6$	 &$260^{+640}_{-190}$	&$15.8^{+0.3}_{-0.9}$ \\
\noalign{\medskip}
\mgxii\ \lya	&1.473	& 8.417  &  1.557 & 7.963 &$-16670^{+630}_{-600}$	
                &$20.9 \pm 12.0$	&$1150^{+1040}_{-870}$	&$17.4^{+0.2}_{-0.4}$\\
\noalign{\smallskip}
\mgxi\ \hea	&1.354	& 9.157  &  1.433 & 8.652  &$-17070^{+450}_{-180}$	
                &$1.7\pm 1.4$	&$85^{+1230}_{-70}$	&$15.5^{+0.3}_{-0.7}$\\
\noalign{\medskip}
\sixiv\ \lya	&2.006	& 6.181  &  2.128  & 5.826  &$-17690^{+180}_{-30}$	
                &$1.1 \pm 1.0$	&$80^{+930}_{-60}$	&$16.4^{+0.3}_{-1.4}$\\
\noalign{\smallskip}
\sixiii\ \hea	&1.867	& 6.641  &  1.973  & 6.284 &$-16550^{+90}_{-570}$	
                &$11.5\pm 6.5$	&$800^{+540}_{-780}$	&$17.3^{+0.2}_{-0.4}$\\
\noalign{\smallskip}
  \noalign{\smallskip}\hline
%\noalign{\smallskip}
\end{tabular}
\end{center}
\begin{tablenotes}
\item[1]\textbf{Notes.} The columns list the line identification,
  laboratory energy, laboratory wavelength, observed energy in the
  rest frame, observed wavelength in the rest frame, measured outflow
  velocity, line equivalent width, estimated turbulent velocity, and
  estimated ionic column of absorption line. All errors are quoted at
  the 90\% confidence level.
\end{tablenotes}
\end{table*}

\subsection{Continuum Modeling}
\label{pg1211:results:continuum}

We used \xspec package v.\,12.9.0 \citep[][]{Arnaud1996} in \isis
\citep[v.\,1.6.2-35;][]{Houck2000} to model the intrinsic continuum of
the combined \meg and \heg spectra in the (rest frame) energy range of
0.54--7.3 keV. We matched the continuum with a
phenomenological model of $\tbnew \times \highecut \times
(\diskbb + \zpowerlw)$. 
The soft excess continuum below 1.1\,keV is dominated by the
\xspec accretion disk model \diskbb
\citep{Mitsuda1984,Makishima1986}, consisting of multiple blackbodies,
that has one physical parameter: the peak temperature ($T_{\rm in}$)
at the apparent inner disk radius ($R_{\rm in}$)
\citep[see][]{Makishima1986,Kubota1998}. The hard excess continuum
above 1.1\,keV is dominated by the \xspec power-law model
\zpowerlw, which describes the Comptonization corona of the
accretion disk with the photon index ($\Gamma$). 
To account for foreground absorption affecting the shape of the
soft X-ray continuum,
we used the \tbnew component \citep{Wilms2000} for better
flexibility in fitting the continuum shape in our phenomenological model.
However, it does not necessarily represent any quantitative physical property
of our Galaxy or the host galaxy of \pg. 
A high-energy exponential cutoff (\highecut with cutoff energy
$E_{c}$ and folding energy $E_{f}$ listed in
Table~\ref{pg1211:fit:continuum}) was also used to account
phenomenologically for the curvature in the hard X-ray spectrum. 
Table~\ref{pg1211:fit:continuum} details the best fit parameters based
on this model.

\begin{figure}
\begin{center}
\includegraphics[width=3.2in, trim = 0 0 0 0, clip, angle=0]{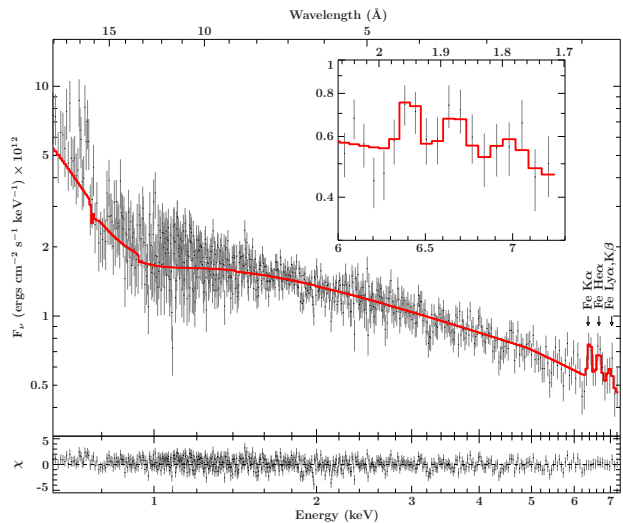}
\caption{The X-ray spectrum taken with the \chandra-\hetgs in 2015
  April averaged over 390\,ks (i.e., excluding the 4th
  observation). The upper panel shows the first-order spectrum (black
  line) and the best-fitting continuum model plus iron emission lines
  (red line; \tbnew*\highecut*(\diskbb+\zpowerlw+$\sum$~\zgauss)),
  while the lower panel plots the $\chi^2$ residuals between the
  observation and the best-fitting model.
\label{fig:pg1211:continuum}%
}
\end{center}
\end{figure}

\subsection{Iron Emission Lines}
\label{pg1211:iron:emissions}

To fit the iron emission lines, we added four Gaussian components
(\zgauss) at 6.4\,keV (Fe\,K$\alpha$), 6.67\,keV (Fe\,He$\alpha$),
6.96\,keV (Fe\,Ly$\alpha$), and 7.05\,keV (Fe\,K$\beta$) in the rest
frame to the continuum model as follows: $\tbnew \times \highecut
\times (\diskbb + \zpowerlw + \sum_{i=1}^{4} \zgauss(i))$.

As seen in Figure \ref{fig:pg1211:continuum}, the data are consistent
with three Gaussian lines at energies between 6\,keV and 7\,keV,
namely the K$\alpha$ fluorescent iron line at 6.4\,keV, the He$\alpha$
iron line at 6.67\,keV, and the Fe Ly$\alpha$ iron line at 6.96 keV
blended with K$\beta$ fluorescent iron lines at 7.05. To measure
 the Ly$\alpha$ line flux correctly, we set the K$\beta$ line flux to
the theoretical iron K$\beta$/K$\alpha$ flux ratio of 0.135
\citep[e.g. see][]{Palmeri2003}, then estimated the Ly$\alpha$ line in
the presence of a possible K$\beta$ contribution, whose absolute width
and relative energy were also tied to the Fe K$\alpha$. 
The lines that we associate with fluorescent iron and He-like iron
emission are significantly detected with positive equivalent widths.

\subsection{Ionized Absorption Lines}
\label{pg1211:iron:absorptions}

Strong absorption lines of key H- and He-like ions were visually identified
initially, and we modeled them using a series of Gaussian absorption functions
(\gabs) with physical properties listed in
Table~\ref{tab:list:absorption}. For thoroughness, we also identified
additional potential spectral features using a blind line search described in
the Appendix.  For these key lines, we obtained the ionic column
densities ($N_{j}$) from the following relationship between
$W_{\lambda}$ and $N_{j}f_{jk}$ for the unsaturated absorption lines
\citep[][Ch. 3]{Spitzer1978}:
\begin{equation}
\frac{W_{\lambda}}{\lambda}=\frac{\pi e^2}{m_e c^2} N_{j} \lambda
f_{jk},\label{eq_3}\\
\end{equation}
where $W_{\lambda}$ is the line equivalent width, $\lambda$ the
wavelength, $N_{j}$ the ionic column density, $f_{jk}$ the oscillator
strength of the relevant transition taken from the atomic database
AtomDB \citep[v.\,2.0.2;][]{Foster2012}, $c$ the speed of light, $m_e$
the electron mass, and $e$ the elementary charge.

Table~\ref{tab:list:absorption} lists the measured outflow velocity
($v_{\rm out}$).
and the line equivalent width ($W_{\lambda}$) in m{\AA}
from the absorption lines. The turbulent velocity width is estimated
from the FWHM ($=2\sqrt{2\ln 2} \sigma$, where $\sigma$ is the velocity
dispersion) of the line profile using $v_{\rm turb}={\rm FWHM}/2
\sqrt{\ln 2}$. The ionic column density is derived using equation
(\ref{eq_3}).

\section{Photoionization Modeling}
\label{pg1211:modeling}
We next proceed with more detailed photoionization analysis using the
code \xstar
\citep[v\,2.2;][]{Kallman1996,Kallman2001,Kallman2004,Kallman2009},
primarily developed for X-ray astronomy. This code solves the
radiative transfer of ionizing radiation in a spherical gas cloud
under a variety of physical conditions for astrophysically abundant
elements, calculates ionization state and thermal balance, and
produces the level populations, ionization structure, thermal
structure, emissivity and opacity of a gas with specified density and
composition, including the rates for line emission and absorption from
bound-bound and bound-free transitions in ions. It utilizes the atomic
database of \citet{Bautista2001}, containing a large quantity of
atomic energy levels, atomic cross section, recombination rate
coefficients, transition probabilities, and excitation rates.

\begin{figure}
\begin{center}
\includegraphics[width=3.4in, trim = 0 0 0 0, clip, angle=0]{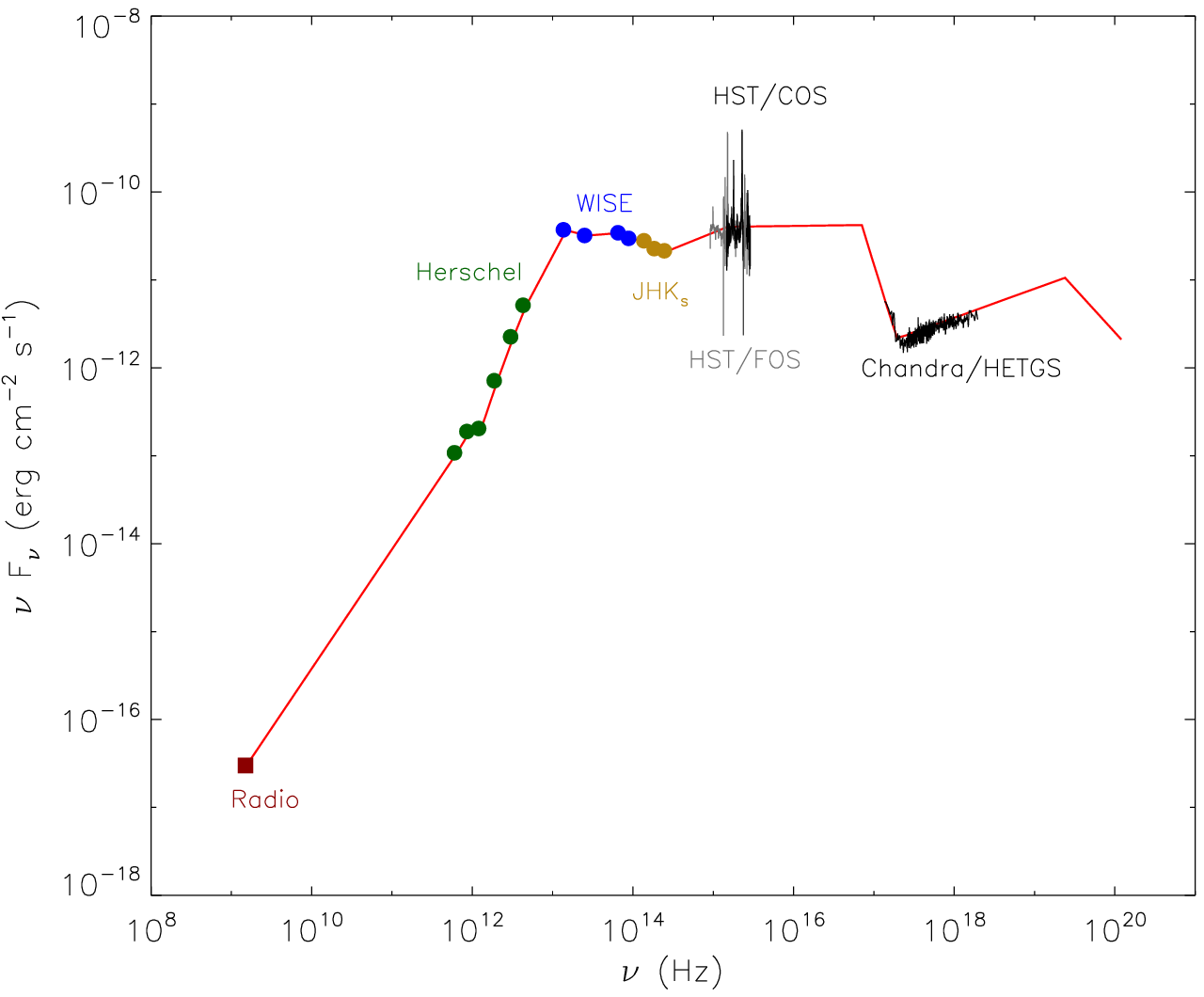}%
\caption{Observed and interpolated baseline SED (solid red line) used
  for photoionization modeling of \pg. The solid points are (1) the
  radio data at 20\,cm taken with the \jvla, and (2) infrared
  measurements (\textit{JHK}s, 3.4, 4.6, 12, 22, 70, 100, 160, 250,
  350, and 500 $\mu$m) from \citet{Petric2015}. Spectra are also shown
  (in solid black and gray lines), and are (1) the archival UV data
  taken with the \hstfos in 1991 April 1(PID 1026, PI: Burbidge), (2)
  our recent FUV observations with the \hstcos, and (3) the
  \chandra-\hetgs.
\label{fig:pg1211:sed}%
}
\end{center}
\end{figure}

We assumed a spherical geometry with a covering fraction of $C_{f}
=\Omega / 4 \pi= 0.5$, which is typical of the ionized absorbers in
AGNs \citep{Tombesi2010a}.  Fundamental parameters in photoionization
modeling are the total gas number density $n$ (in cm$^{-3}$), the
total hydrogen column density $N_{\rm H}= n_{H} V_{f} \triangle r$ (in
cm$^{-2}$), the ionization parameter $\xi=L_{\rm ion}/n_{H} r^2$
\citep[in erg\,cm\,s$^{-1}$;][]{Tarter1969}, and the turbulent
velocity $v_{\rm turb}$ (km\,s$^{-1}$), where $L_{\rm ion}$ (in
erg\,s$^{-1}$) is the ionizing luminosity between 13.6 eV and 13.6 keV
(i.e. 1 and 1000 Ryd), $V_{f}$ is the volume filling factor of the
ionized gas, and $r$ and $\triangle r$ are the distance from the
ionizing source and the thickness of the ionized gaseous shell (in
units of cm), respectively. We set the initial gas temperature to the
typical value of $T_{\rm init}= 10^{6}$\,K
\citep{Nicastro1999,Bianchi2005}, and allowed the code to calculate it
based on the thermal equilibrium of gas. We generated grids of
\xstar models (see \S\,\ref{pg1211:results:sed:grids}) using
an ionizing spectral energy distribution (SED) described in
\S\,\ref{pg1211:results:sed} (see also Figure\,\ref{fig:pg1211:sed}).

\subsection{Spectral Energy Distribution}
\label{pg1211:results:sed}

Since it was found by \citet{Lee2013} that the UV continuum is an
important component of the ionizing flux, we employed a similar
methodology here to generate radio, infrared, UV, and X-ray components
of the ionizing SED for the photoionization modeling of the warm
absorber in \pg.  Although the combined UV and X-ray continua are
expected to be the main source of the ionizing radiation, we have used
the entire ranges from radio to X-ray to construct the SED (see
Figure\,\ref{fig:pg1211:sed}).  The archival UV data taken with the
\hstfos in April 1991 (G130H, G190H, and G270H), together
with our \hstcos time-averaged FUV spectrum (G140L) observed
simultaneously with \chandra in April 2015, were used to
construct the SED. 
We utilized our \chandra\ \meg and \heg data to make X-ray
continuum regions (0.5--8\,keV) of the ionizing SED.  The soft X-ray
continuum was extrapolated as a power law at energies below 0.5 keV to
the point at which it meets the high-energy extrapolation of the UV
powerlaw. 
We also used the radio fluxes at 20\,cm (1.5\,GHz), $S_{\nu}=2$ mJy,
measured with the \jvla (\S~\ref{pg1211:observations:radio}), the near-infrared (NIR)
measurements (\textit{JHK}s) from the Two Micron All Sky Survey
(2MASS), the mid-infrared (MIR) measurements at 3.4, 4.6, 12, and 22
$\mu$m from the Wide-field Infrared Survey Explorer (WISE), and the
far-infrared (FIR) measurements at 70, 100, 160, 250, 350, and 500
$\mu$m from the ESA \herschel
\citep{Petric2015}. Similarly, the radio, NIR, MIR, and FIR band
points were connected to each other. However, the ionizing SED is
mainly characterized by the UV and X-ray spectra without the emission
and absorption lines.  The IR, optical and UV data were first dereddened
using $R_{\rm V}= 3.1$ and $E({\rm B-V})= 0.035$ \citep{Schlafly2011},
and placed in the rest frame. The X-ray data were also corrected
for the foreground Galactic absorption.

The resulting intrinsic SED is shown in Figure \ref{fig:pg1211:sed}
with associated bands (points) and composite spectra,
connecting the IR, UV and X-rays regions (solid line). The
intrinsic SED is then used to generate grids of photoionization models
that are fitted to the X-ray absorption lines. To obtain the ionizing
luminosity, we integrate the interpolated baseline SED between
$\nu=3.29 \times 10^{15}$ and $3.29 \times 10^{18}$ Hz (i.e., 1--1000
Ryd), finding $\int F_{\nu}\,d\nu = 1.035 \times 10^{-10}$
erg\,cm$^{-2}$\,s$^{-1}$, which yields $L_{\rm ion} =1.587 \times
10^{45}$ erg\,s$^{-1}$ at the luminosity distance of 358\,Mpc ($H_{0}=
73$\,km\,s$^{-1}$\,Mpc$^{-1}$, $\Omega_{m} = 0.27$, and $\Lambda_{0} =
0.73$; $z=0.082$ corrected to the microwave background radiation
reference frame). Previously, the ionizing luminosity (1--1000 Ryd) of $3.8 \times
10^{45}$ erg\,s$^{-1}$ was estimated from XMM--Newton observation \citep{Pounds2016b}.

\subsection{Photoionization Tabulated Grids}
\label{pg1211:results:sed:grids}

\begin{table}
\caption{Parameter ranges for \xstar Photoionization Model Grids.
\label{pg1211:fit:input}
}
\centering
\begin{tabular}{C{4.4cm}cc}
\hline\hline
\noalign{\smallskip}
{Parameter} & {Value}  & {Interval Size}  \\
\noalign{\smallskip}
\hline 
\noalign{\smallskip}
$L_{\rm ion}$ ($10^{38}$ erg\,s$^{-1}$)\dotfill  & $1.587\times 10^{7}$  & --\\ 
\noalign{\smallskip}
$T_{\rm init}$ ($10^{4}$ K)\dotfill  & $ 100$   &  --\\ 
\noalign{\smallskip}
$\log n$ (cm$^{-3}$)\dotfill  & $8 \cdots 14$   & $1.0$\\ 
\noalign{\smallskip}
$\log N_{\rm H}$ (cm$^{-3}$)\dotfill  & $18 \cdots 25$   & $0.5$\\ 
\noalign{\smallskip}
$\log \xi$ (erg\,cm\,s$^{-1}$)\dotfill  & $-2 \cdots 5$  & $0.25$ \\
\noalign{\smallskip}
$v_{\rm turb}$ (km\,s$^{-1}$)\dotfill  & $100 \cdots 500$  & $100.0$\\ 
\noalign{\smallskip}
$A_{\rm Fe}$ \dotfill  & $ 1.0$   & --\\ 
\noalign{\smallskip}
$C_{f}=\Omega / 4 \pi$\dotfill  & $ 0.5$   & --\\ 
\noalign{\smallskip} 
\hline
\noalign{\smallskip}
\end{tabular}
\begin{tablenotes}
\item[1]\textbf{Notes.} Logarithmic interval sizes are chosen for the
  total gas number density ($n$), the total hydrogen column density
  ($N_{\rm H}$), the ionization parameter ($\xi=L_{\rm ion}/n_{H}
  r^2$), while a linear interval size is chosen for the turbulent
  velocity ($v_{\rm turb}$).  The ionizing luminosity ($L_{\rm ion}$)
  is derived from an integration (1--1000 Ryd) of the interpolated
  baseline SED shown in Figure~\ref{fig:pg1211:sed}. The initial gas
  temperature ($T_{\rm init}$) is set to the typical value of
  $10^6$\,K \citep[][]{Nicastro1999,Bianchi2005}, while the code
  adjusts it based on the thermal equilibrium. The covering fraction
  is fixed to the typical value of the X-ray absorbers
  \citep[$C_{f}=0.5$;][]{Tombesi2010a}.
\end{tablenotes}
\end{table}

\begin{figure*}
\begin{center}
\includegraphics[width=6.3in, trim = 0 0 0 0, clip, angle=0]{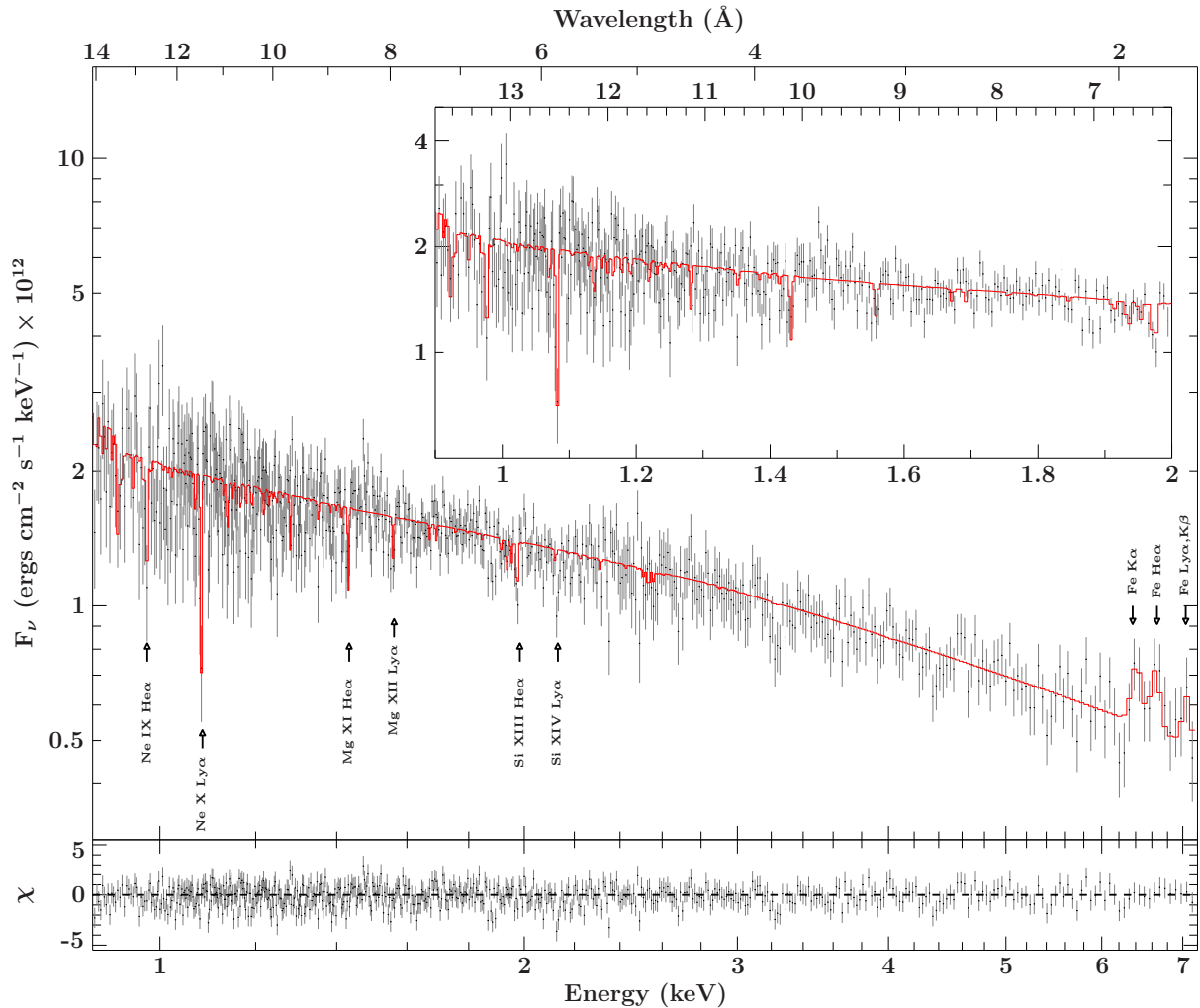}
\caption{\chandra\ \hetgs spectrum for \pg fit with the \xstar \warmabs model (see
  Table \ref{pg1211:fit:absorbers}).  The lower panel plots the
  $\chi^2$ residuals between the observation and the model.
\label{fig:pg1211:spectrum}%
}
\end{center}
\end{figure*}

We used the ionizing SED made in \S\,\ref{pg1211:results:sed} to
generate grids of \xstar models. We utilized
\mpixstar\footnote{\url{http://github.com/xstarkit/mpi\_xstar/}} \citep{Danehkar2018} that
allows parallel execution of multiple \xstar runs on a computer
cluster (in this case, the \odyssey cluster at Harvard University). 
It employs the \textsf{xstar2table} script (v.\,1.0) to produce multiplicative tabulated model files: an absorption spectrum imprinted onto a continuum (\textsf{xout\_mtable.fits}), a reflected emission spectrum in all directions (\textsf{xout\_ain.fits}), and an emission spectrum in the transmitted direction of the absorption (\textsf{xout\_aout.fits}).  The first and second tabulated model files are used as absorption and emission components of the ionized outflows (or infalls) in spectroscopic analysis tools. 
Table~\ref{pg1211:fit:input} lists the parameters used for
producing \xstar model grids. To cover the possible range of physical
conditions, we initially considered a large range of gas densities $n$
from $10^{8}$ to $10^{14}$\,cm$^{-3}$, column densities $N_{\rm H}$
from $10^{18}$ to $10^{25}$\,cm$^{-2}$, ionization parameters $\xi$
from $10^{-2}$ to $10^{5}$ erg\,cm\,s$^{-1}$, and turbulent velocities
of 100--500 km\,s$^{-1}$, for use in spectral fitting. 

We computed a grid of $15 \times 29$ \xstar models on the
two-dimensional $N_{\rm H}$--$\xi$ plane, sampling the fundamental
parameter space with 15 logarithmic intervals in the column density
(from $\log N_{\rm H}=18$ to $25$\,cm$^{-2}$ with the interval size of
$0.5$) and 29 logarithmic intervals in the ionization parameter (from
$\log\xi=-2$ to $5$ erg\,cm\,s$^{-1}$ with the interval size of
$0.25$), assuming a gas density of $n=10^{12}$\,cm$^{-3}$.  For
diagnostic purposes, we have initially constructed some grids with the
same $N_{\rm H}$--$\xi$ parameter space for $\log
n=8$--$14$\,cm$^{-3}$ (interval size of $1$), and $v_{\rm
  turb}=100$--$500$\,km\,s$^{-1}$ (interval size of $100$).  However,
we found that results were indistinguishable for this wide range of
the gas density in highly-ionized absorbers.  Hereafter, all
\xstar models correspond to a gas density of
$n=10^{12}$\,cm$^{-3}$.

The turbulent velocity is another important parameter in
photoionization modeling. An increase in the turbulent velocity
increases the equivalent width of an absorption line for a given
column density of each ion. To estimate equivalent widths correctly,
the velocity width, which is associated with line broadening, must be
measured precisely. The high spectral resolution of \meg and \heg data
allows for the possibility of determining the velocity width.  However, 
due to insufficient counts, our \hetgs observations pointed to a wide
range of velocity widths with high uncertainties, so we could not
measure the exact value of $v_{\rm turb}$ for each ion. 
From the line width measurements (see
Table~\ref{tab:list:absorption}), we adopted a velocity turbulence of
$v_{\rm turb}=200$\,km\,s$^{-1}$ for the warm absorber, which
approximately corresponds to the \hetgs optimal spectral resolution. 

\begin{figure*}
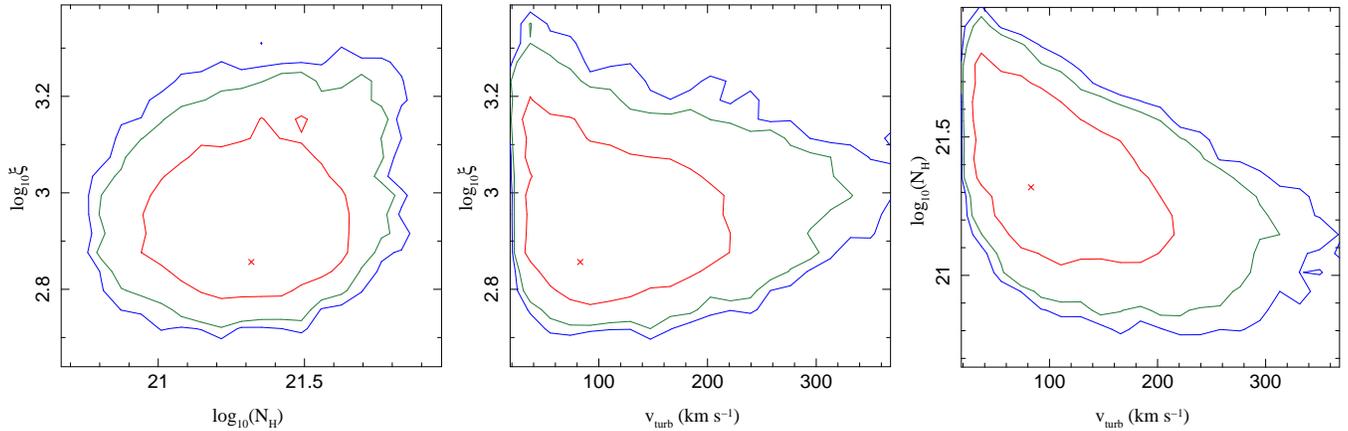

\begin{center}
\includegraphics[width=2.35in, trim = 30 30 0 0, clip, angle=0]{figures/fig8_conf_xi_nh.ps}% 
\includegraphics[width=2.35in, trim = 30 30 0 0, clip, angle=0]{figures/fig8_conf_xi_vturb.ps}%
\includegraphics[width=2.35in, trim = 30 30 0 0, clip, angle=0]{figures/fig8_conf_nh_vturb.ps}%
\caption{The 1-sigma (68\%), 2-sigma (95\%), and 3-sigma (99\%)
  confidence contours of the logarithm of the ionization parameter
  ($\log\xi$) vs. the logarithm of the column density ($\log N_{\rm
    H}$; left), the logarithm of the ionization parameter ($\log\xi$)
  vs. the turbulent velocity ($v_{\rm turb}$; middle), and the
  logarithm of the column density ($\log N_{\rm H}$) vs. the turbulent
  velocity ($v_{\rm turb}$; right) of the ionized absorber with the
  best-fitting parameters listed in Table~\ref{pg1211:fit:absorbers},
  respectively.  The cross sign shows the best-fitting values in each
  panel.
\label{fig:pg1211:confmap}%
}
\end{center}
\end{figure*}

We proceeded to fit the combined \meg and \heg data shown in
Figure~\ref{fig:pg1211:spectrum}, multiplying our continuum model by 
the \xstar tabulated grids produced from the ionizing SEDs in \S\,\ref{pg1211:results:sed}. 
There are a total of two free parameters in the
tabulated grid fitting, namely, the ionization parameter $\xi$ and
column density $N_{\rm H}$ of the ionized absorber. Using the base
continuum model described in \S\,\ref{pg1211:results:continuum}, the
model for the spectra containing 1 ionized absorber and 4 Fe emission
lines are implemented as follows: $\tbnew \times \highecut \times
(\diskbb + \zpowerlw + \sum_{i=1}^{4} \zgauss(i)) \times \xstarabs$. 
These models successfully described the data with an ionization
parameter of $\log \xi=2.9\pm0.1$ and a column density of $N_{\rm
  H}=(2.4\pm0.6)\times10^{21}$ (90\% confidence levels), with an observed
redshift of $(0.0201\pm0.0005)c$. The model with these parameters has a goodness-of-fit of $\chi^{2}/{\rm d.o.f}=738/661$. 
These results allowed us to refine our parameter estimations in
performing the \warmabs model fitting described in the subsequent
section.

\subsection{Photoionization Analytic Modeling}
\label{pg1211:results:sed:fitting}

We constrained the properties of potential ionized absorbers using the
\xstar analytic model \warmabs, relying on all atomic level
populations calculated by \xstar using the same UV--X-ray ionizing SED
from \S\,\ref{pg1211:results:sed}. Fitting the spectrum with the
tabulated grids (see \S\,\ref{pg1211:results:sed:grids}) is much
quicker than the \warmabs analytic modeling. However, parametric
intervals used in the computation of the model grids (see interval
sizes in Table~\ref{pg1211:fit:input}) could add large uncertainties
to derived parameters. Hence, we also used the analytic model to put
more accurate constraints on the ionization state and column density
of the warm absorbing gas, as well as to explore any variations in
elemental abundances. As this approach is computationally expensive,
we utilized the parameters determined from the \xstar table model
grids in \S\,\ref{pg1211:results:sed:grids} as initial values, and
employed \warmabs only to make finer estimations of the physical
conditions.

For the soft excess, we again use an absorbed disk component.
The high-ionization lines visible in the spectrum
({\it c.f.} Table \ref{tab:list:absorption}) clearly
require a highly ionized \warmabs component.
Given that the \tbnew component in our phenomenological model (\S 4.2) shows
more absorption than expected for foreground Milky Way absorption 
\citep[$N_{\rm H} = 2.6 \times 10^{20}$~cm$^{-2}$;][]{Wakker2011},
we have tried three different options for additional absorption in \pg.
Specifically, we considered (1) a neutral absorber (\tbnew) at the \pg
systemic redshift,
(2) a mildly ionized \warmabs model (also at the \pg\  systemic redshift), and
(3) a low-ionization \warmabs model at the same observed redshift as the
high-ionization \warmabs component.\footnote{Lacking compelling statistics
  for any redshift other than the component with
  $z_{\rm obs}\approx0.02$, we choose this value to keep the fit tractable.}
For this third possibility we also
tied the abundance and turbulent velocity parameters to those of the
high-ionization \warmabs component.

\begin{table}
\caption{Best-fitting parameters for the \xstar \warmabs model, 
the continuum model and iron lines obtained using the \isis 
\emcee routine.
\label{pg1211:fit:absorbers}
}
\centering
\begin{tabular}{lC{3.3cm}c}
\hline\hline
\noalign{\smallskip}
{Component} & {Parameter} & {Value}    \\
\noalign{\smallskip}
\hline 
\noalign{\smallskip}
 & $\chi^{2}/{\rm d.o.f}$ \dotfill &  $610/614$ \\ 
\noalign{\smallskip}
\highecut & $E_{c}$(keV) \dotfill & $1.4^{+0.3}_{-0.4}$   \\ 
\noalign{\smallskip}
          & $E_{f}$(keV) \dotfill & $11.3^{+3.5}_{-0.8}$    \\ 
\noalign{\medskip}                   
\diskbb & $T_{\rm in}$(keV) \dotfill & $0.096^{+0.002}_{-0.010}$  \\ 
\noalign{\smallskip}
        & \textit{Norm} \dotfill  & $1.2^{+1.5}_{-0.3}\times 10^{4}$   \\ 
\noalign{\medskip} 
\zpowerlw & $\Gamma$ \dotfill  & $1.55^{+0.04}_{-0.07}$   \\ 
\noalign{\smallskip}
        & \textit{Norm} ($\gamma$\,keV$^{-1}$\,cm$^{-2}$) \dotfill 
        & $1.38^{+0.04}_{-0.09} \times 10^{-3}$    \\ 
\noalign{\medskip} 
\textsf{zgauss$_{\rm K\alpha}$} & $E$(keV) \dotfill & $6.41^{+0.06}_{-0.05}$   \\ 
\noalign{\smallskip}
(Fe\,K$\alpha$)               & $\sigma$(keV) \dotfill & $0.004^{+0.13}_{}$   \\ 
\noalign{\smallskip}
        & \textit{Norm} ($\gamma$\,cm$^{-2}$\,s$^{-1}$) \dotfill 
        & $3.6^{+3.2}_{-2.0}\times10^{-6}$ \\ 
\noalign{\medskip}       
\textsf{zgauss$_{\rm He\alpha}$} & $E$(keV) \dotfill & $6.66^{+0.12}_{-0.09}$   \\ 
\noalign{\smallskip}
(\fexxv)                       & $\sigma$(keV) \dotfill & $0.06^{+0.22}_{-0.02}$   \\ 
\noalign{\smallskip}
                & \textit{Norm} ($\gamma$\,cm$^{-2}$\,s$^{-1}$) \dotfill  
                & $5.2^{+6.6}_{-3.9}\times10^{-6}$ \\  
\noalign{\medskip} 
\warmabs  & $\log n$ (cm$^{-3}$) \dotfill  & 12.0   \\ 
\noalign{\smallskip}
          & $\log N_{\rm H}$ (cm$^{-2}$) \dotfill  &  $21.47^{+0.18}_{-0.58}$    \\ 
\noalign{\smallskip}
          & $\log \xi$ (erg\,cm\,s$^{-1}$) \dotfill  &  $2.87^{+0.32}_{-0.10}$    \\ 
\noalign{\smallskip}
          & $v_{\rm out}$ (km\,s$^{-1}$) \dotfill  &  $-17300^{+100}_{-130}$    \\ 
\noalign{\smallskip}
          & $v_{\rm turb}$  (km\,s$^{-1}$)  \dotfill  &  $ 90^{+210}_{-60}$    \\ 
\noalign{\medskip} 
          & $A_{\rm Ne}$  \dotfill  &  $0.6^{+2.0}_{-0.2}$    \\ 
\noalign{\smallskip}
          & $A_{\rm Mg}$  \dotfill  &  $1.5^{+3.2}_{-0.4}$    \\ 
\noalign{\smallskip}
          & $A_{\rm Si}$ \dotfill  &  $3.0^{+4.8}_{-1.1}$    \\ 
\noalign{\smallskip}
          & $A_{\rm S}$ \dotfill  &  $0.6^{+5.3}_{-0.4}$    \\ 
\noalign{\smallskip}
          & $A_{\rm Fe}$ \dotfill  &  $1.2^{+1.6}_{-0.6}$    \\ 
\noalign{\smallskip}
\warmabs  & $\log N_{\rm H}$ (cm$^{-2}$) \dotfill  &  $21.08^{+0.10}_{-0.62}$    \\ 
\noalign{\smallskip}
          & $\log \xi$ (erg\,cm\,s$^{-1}$) \dotfill  &  $1.29^{+0.34}_{-0.29}$    \\ 
\noalign{\smallskip} 
\hline
\noalign{\smallskip}
\end{tabular}
\begin{tablenotes}
\item[1]\textbf{Notes.}  The total gas number density ($n$) is fixed
  to the typical value for AGN ionized absorbers. The total hydrogen
  column density ($N_{\rm H}$), ionization parameter ($\xi=L_{\rm
    ion}/n_{H} r^2$), Doppler velocity ($v_{\rm out}$),
  turbulent velocity ($v_{\rm turb}$), and elemental abundances ($A$)
  are free parameters estimated by the \xstar analytic \warmabs
  modeling.  The chemical abundances are with respect to the solar
  values defined in \citet{Wilms2000}. The second \warmabs component
  has all parameters, save ionization parameter and column, tied to
  those of the more highly ionized absorber. All uncertainties are
  estimated at the 90\% confidence level.
\end{tablenotes}
\end{table}

All three of these possibilities resulted in essentially identical parameters
for the highly ionized absorber. None of the three possibilities produced
identifiable spectral features, but rather affected the shape of the
low-energy X-ray continuum. Differences among the fits were primarily
restricted to the soft excess parameters, with subtler differences in the
highest and \zpowerlw parameters.
In what follows, we only present results for option (3), the additional
low-ionization component at the observed redshift of the high-ionization
component.
However, we consider the high-ionization component to be the only one with
well-measured parameters, and we only discuss its physical implications.

In our fits, we used the solar abundances defined by
\citet{Wilms2000}, and adjusted the elemental abundances and the
turbulent velocity to match the absorption lines. The
best-fitting parameters derived from our model fits
are listed in Table\,\ref{pg1211:fit:absorbers}. 
We determined the fit parameter confidence intervals 
by employing the \isis \emcee routine, which is an implementation of
the Markov Chain Monte Carlo (MCMC) Hammer algorithm of
\citet{Foreman-Mackey2013}, and now included in the Remeis
ISISscripts\footnote{\url{http://www.sternwarte.uni-erlangen.de/isis/}}. The
MCMC approach allowed us to improve the best-fitted values of the
ionization parameter ($\xi$), the column density ($N_{\rm H}$), the
turbulent velocity ($v_{\rm turb}$), and the chemical composition of the
\warmabs model as well as other phenomenological model parameters
(continuum and iron lines) listed in Table~\ref{pg1211:fit:absorbers} 
(the fits, however, are not very sensitive to the chemical
composition). Figure~\ref{fig:pg1211:confmap} shows the 1-sigma
(68\%), 2-sigma (95\%), and 3-sigma (99\%) confidence contours of
$\log\xi$ versus $\log N_{\rm H}$, $\log\xi$ versus $v_{\rm turb}$,
and $\log N_{\rm H}$ versus $v_{\rm turb}$.  There is a slight
anti-correlation between the absorber column and turbulent velocity
(somewhat expected to fit the equivalent widths of the detected
features).  Overall, the main absorber parameters are
well constrained. 

\begin{figure}
\begin{center}
\includegraphics[width=3.3in, trim = 0 0 0 0, clip, angle=0]{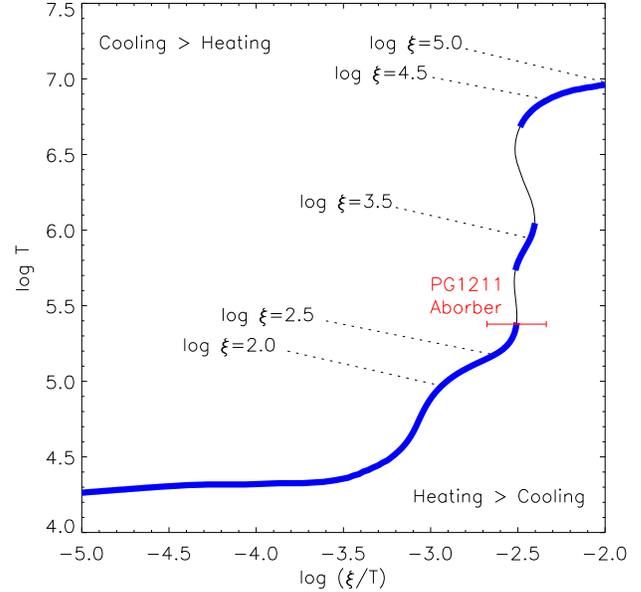}%
\caption{Thermal stability curve for $n_{\rm H} = 10^{12}$\,cm$^{-3}$
  produced using the \xstar model associated with the UV--X-ray
  ionizing SED shown in Figure\,\ref{fig:pg1211:sed}.  It shows the
  distribution of equilibrium temperature $\log T$ as a function of
  $\log (\xi/T)$.  The thick solid lines in the curves correspond to
  the regions with the thermally stable gas.  The best-fitting $\xi$
  value of the ionized absorber component is labeled as PG1211 with
  its corresponding error bar at the 90\% confidence level. 
\label{fig:pg1211:stability}%
}
\end{center}
\end{figure}

\subsection{Thermal Stability}
\label{pg1211:results:stability}

The photoionization model fitting yielded the ionized absorber with
the physical conditions listed in Table~\ref{pg1211:fit:absorbers},
hydrogen-equivalent column density $N_{\rm H}$ (in cm$^{-2}$) and
ionization parameter $\log \xi$ (in erg\,cm\,s$^{-1}$), which are
required to reproduce the blueshifted absorption features in the
spectrum of \pg.

\begin{figure}
\begin{center}
\includegraphics[width=3.2in, trim = 50 30 0 0, clip, angle=0]{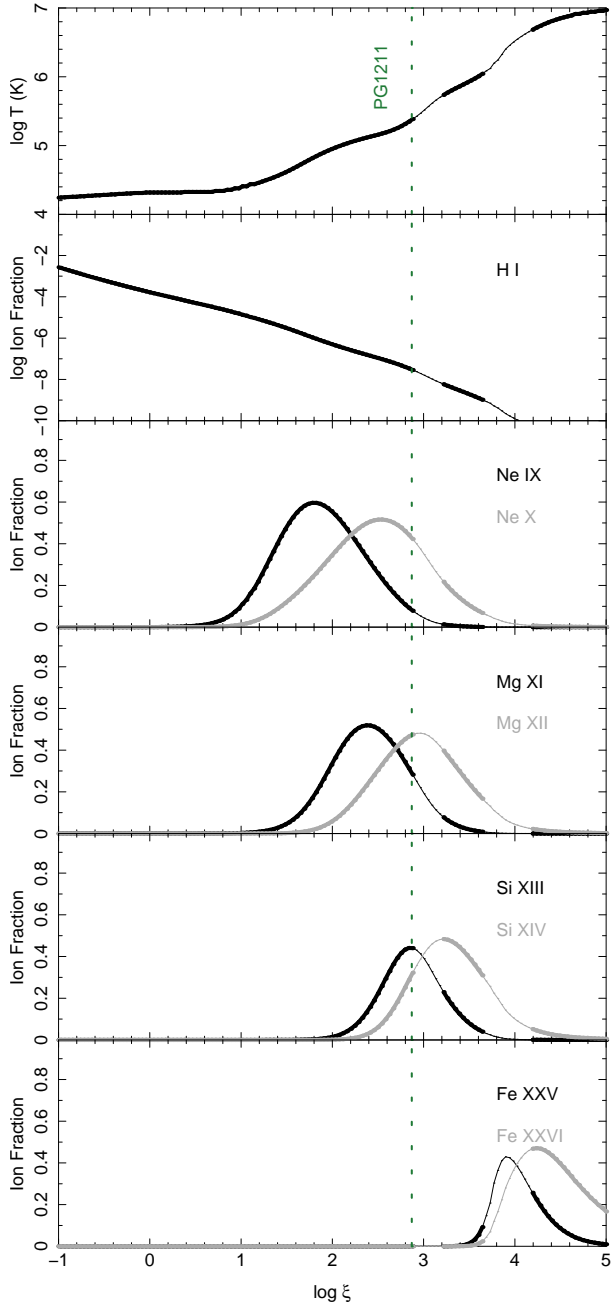}%
\caption{ The distribution of temperature $\log T$ (top panel) and the
  ion fraction distributions of the neutral hydrogen (\hi), the
  He-like (black line) and H-like (gray line) ions of the relevant
  elements (Ne, Mg, Si and Fe) as a function of ionization parameter
  $\log \xi$ produced using the \xstar model with the ionizing
  UV--X-ray SED shown in Figure\,\ref{fig:pg1211:sed}.  The thick solid
  lines in the curves correspond to the regions with the thermally
  stable gas.  The vertical dot line (labeled as PG1211) is associated
  with the \xstar warm absorber with parameters listed in
  Table~\ref{pg1211:fit:absorbers}.  We note that while \fexxv\, is
  included here, it was not detected with any significance in this
  {\it Chandra} observation.  The gas density is chosen to be $n_{\rm H} =
  10^{12}$\,cm$^{-3}$.
As can be seen, H~{\sc i} can exist even at 
the high ionization required by the detected X-ray lines,
lending extra credence to 
our findings that the $\rm \sim 16980 \,km \, s^{-1}$ outflow we detect with 
\chandra and \hst are likely associated.
\label{fig:pg1211:ionfrac}%
}
\end{center}
\end{figure}

The stability curve, in which temperatures ($T$) of clouds are plotted
against their pressures ($\xi/T$), is an effective theoretical tool to
illustrate the thermal stability of ionized absorbing clouds
\citep{Krolik1981,Reynolds1995,Krolik2001,Chakravorty2009,Chakravorty2013}. 
The absorber is thermally stable where the slope of the
stability curve is positive and where the heating and cooling
mechanisms are in equilibrium.  Figure~\ref{fig:pg1211:stability}
shows the stability curve generated using the \xstar model for
a gas density $n_{\rm H} = 10^{12}$\,cm$^{-3}$ and the corresponding
parameters specified in \S\,\ref{pg1211:results:sed:fitting}.  As can be seen,
the ionized absorber is just at the edge of the thermally stable
region. 
Interestingly, in the prior {\it XMM-Newton} observation in 2014, when
PG\,1211+143 was twice as bright, the
ionization parameter $\xi$=3.4 \citep{Pounds2016b},
consistent with the increased brightness, and lying on the next-highest
stable portion of the curve.

Figure~\ref{fig:pg1211:ionfrac} shows the distribution of temperature
$\log T$, the neutral hydrogen (\hi), and ion fractions of the He-like and H-like ions of the
relevant elements (Ne, Mg, Si and Fe) with respect to the ionization
parameter $\log \xi$, from our \xstar
model. The distribution of temperature and ion fractions typically
depend on the ionization parameter, the gas density, and the ionizing
SED \citep{Kallman2001}.  The thick solid lines correspond to the
range where the absorbing gas is thermally stable.  This figure
further shows that both the X-ray detected ions and the UV absorber
detected in \hi\, can all coexist in a single ionization zone at the
same velocity.

\section{\hstcos Results: Evidence for a Corresponding UV Absorber}
\label{pg1211:results:uv}

\begin{figure}
\begin{center}
\includegraphics[width=2.6in, trim = 0 0 0 0, clip, angle=-90]{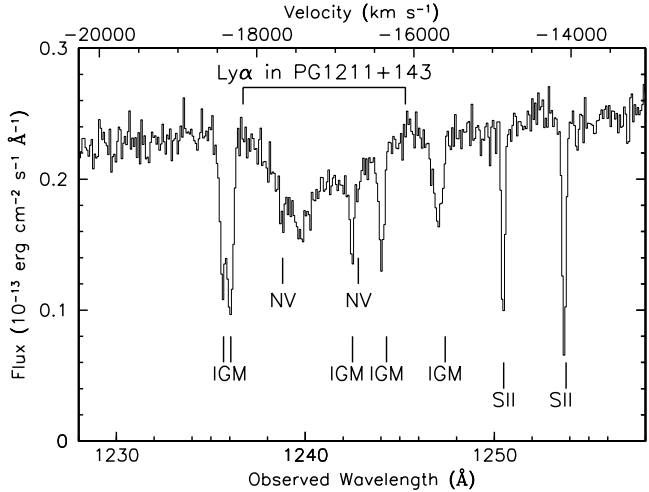}%
\caption{ \hstcos G130M spectrum of \pg in the wavelength range covering
  the broad Ly$\alpha$ absorption line profile, the IGM Ly$\alpha$ lines
  \citep{Penton2004} and interstellar \nv\ and \sii\ lines, adapted from \cite{Kriss2017}. The lower horizontal axis is the observed wavelength
in \AA ngstrom.
The upper horizontal axis is the outflow velocity of Ly$\alpha$ relative to a
systemic redshift of $z = 0.0809$.
\label{fig:pg1211:uv:absorber}%
}
\end{center}
\end{figure}

Our \hstcos observations reveal a previously unknown, weak, broad,
blueshifted Ly$\alpha$ absorption feature consisting of two blended
components, as shown in Figure~\ref{fig:pg1211:uv:absorber}.
\citet{Kriss2017} discuss in
detail the analysis of the UV data in general and this feature
specifically.  They find an average outflow velocity of
$v_{\rm out} = -16\,980 \pm 40~\rm km~s^{-1}$ ($z_{\rm out} = -0.0551$)
with a FWHM of
$1080 \pm 80~\rm km~s^{-1}$.
Both the \hst-detected outflow velocity and width
are consistent with the strongest X-ray absorption features, giving
additional credence to the reality of such a high velocity outflow.
In addition to the Ly$\alpha$ feature in the G130M spectrum, we also
detect Ly$\beta$ at the same velocity in the G140L spectrum, albeit at
lower significance. No other UV ions are detected, including
high-ionization species such as C\,{\sc iv}, N\,{\sc v} or O\,{\sc vi}.

Several other narrow absorption lines appear in the UV spectrum of \pg
embedded within the profile of the newly detected broad Ly$\alpha$
absorption feature. These include foreground interstellar absorption
in the N\,{\sc v} doublet, as well as the previously known features
from the foreground intergalactic medium (IGM) \citep{Tilton2012}.
We can safely conclude that these narrow features are not associated
with the outflow from \pg because they do not share any of
the characteristics typical of absorption lines associated with AGN
outflows-- they are not variable in strength, and they fully cover the
source. The Ly$\alpha$ to Ly$\beta$ optical depth ratios
\citep{Danforth2008,Tilton2012} are consistent with their Doppler
widths, which at $\sim 35$~km~s$^{-1}$ are typical of other IGM
Ly$\alpha$ absorbers.

The similar depths of the broad Ly$\alpha$ and Ly$\beta$ features in
our spectra indicate that they are saturated, so that the UV absorber
only partially covers the continuum source.  Using the depth at the
center of the Ly$\alpha$ absorption feature, we measure a covering
fraction of $C_f = 0.30 \pm 0.04$.  Since the lines are saturated, we
cannot measure an accurate column density, but we can set a lower
limit assuming a covering factor of unity and integrating across the
apparent optical depth of the line profile
\citep[e.g.,][]{Hamann1997}.  This gives $\log N_{\text{\hi}} \gtrsim
14.5$~cm$^{-2}$.  

Despite the high ionization of the X-ray absorber detected at the same
outflow velocity, our photoionization models predict a small residual
column density of neutral hydrogen. At our best-fit $\log \xi = 2.87$,
the ionization fraction of H\,{\sc i} is $2.99_{-2.40}^{+1.49} \times
10^{-8}$.  For our best-fit total hydrogen column density of $\log
N_{\rm H} = 21.47_{-0.58}^{+0.18}$~cm$^{-2}$, we predict $\log
N_{\text{\hi}} = 13.95_{ }^{+0.26}$~cm$^{-2}$, roughly consistent
with our measured lower limit of $\log N_{\text{\hi}} \gtrsim
14.5$~cm$^{-2}$.
As \cite{Kriss2017} show, the Ly$\alpha$ profile actually consists of two
blended components, so it is possible that the X-ray UFO is associated with only
a portion of the Ly$\alpha$ absorption profile.
This would also explain the narrow turbulent velocity we find for the X-ray
absorber compared to the Ly$\alpha$ absorber.

All other UV ions are predicted to have column
densities at least three orders of magnitude lower, rendering them
undetectable even in our high S/N \hstcos spectra.

\section{VLA Results: Evidence for a Jet?}
\label{pg1211:jet}

The radio emission from PG 1211+143 is compact at all the frequencies
and resolutions we have studied. In our deep K-band image, fitting a
Gaussian to the source gives a best-fitting major axis of 22 mas,
corresponding to 33 pc at the distance of the quasar assuming the
cosmological parameters from the previous section. Given that residual
phase uncertainties can induce blurring in a point-like source, this
should probably be taken as an upper limit on the size of the source.

The flux densities measured from the various (non-simultaneous) radio
data available to us are summarized in Table \ref{tab:radiofluxes}. In
all cases these are the integrated flux densities of a Gaussian fitted
to the data around the position of the quasar using the AIPS task JMFIT.

\cite{Kellermann1994} report a 5-GHz flux density of $0.80 \pm 0.1$ mJy
for the source in observations taken in 1983, so, taking this with our
two C-band flux densities, there is some modest evidence for a slow
evolution of the apparent luminosity of the source at the 10\% level
on a timescale of decades. We have no information on whether the fast
variability of the source seen in the X-rays is also present in the
radio.

The radio SED of the source is interesting, because, assuming that
variability can be neglected, we see a spectral index $\alpha$
(defined in the sense $S \propto \nu^{\alpha}$) of $-0.49 \pm 0.06$
between L and C-band, $-0.79 \pm 0.03$ between C and K-band, and $-1.0
\pm 0.4$ internal to the K-band (using the `lower' and `upper' halves
of the band). Thus all the SED measurements are consistent with coming
from optically thin synchrotron emission, with a steeper spectrum at
higher frequencies presumably due to radiative losses. If we further
assume that all the emission we see comes from a region of less than
30 pc in size, this gives a consistent picture in which the radio
emission comes from an optically thin region: for example, assuming a
prolate ellipsoidal emission region with major axis 22 mas and minor
axis half the major axis, and taking $E_{\rm min}$ for the
relativistic electrons to be 5 MeV, the minimum energy in the
synchrotron-emitting particles and field would be around $5 \times
10^{53}$ erg, the equipartition field strength would be of order 700
$\mu$G, and self-absorption would be expected to become important at
frequencies below a few hundred MHz. The source would need to be a
factor of a few smaller before self-absorption would be expected to
affect the lowest frequencies we observe. If the source is strongly
projected, as might be expected for an object that appears as a
quasar, then these constraints are relaxed, since the true physical
size is $\la 30/\sin \theta$ pc where $\theta$ is the angle to the line of
sight.

\begin{table}
  \caption{Radio flux densities for PG\,1211+143}
  \begin{tabular}{rrrr}
\hline
\hline
\noalign{\smallskip}
    Band&Central freq.&Date&Flux density\\
    &(GHz)&&(mJy)\\
\hline
\noalign{\smallskip}
    L (FIRST\,$^{\rm a}$)&1.4&$1999$&$2.10 \pm 0.17$\\
\noalign{\smallskip}
    C&4.9&1993 May 02&$0.95 \pm 0.07$\\
\noalign{\smallskip}
    C&6&2015 Jun 18&$1.032 \pm 0.015$\\
\noalign{\smallskip}
    K&22&2015 Jun 18&$0.370 \pm 0.011$\\
\noalign{\smallskip}
    K (lower)&20&2015 Jun 18&$0.384 \pm 0.013$\\
\noalign{\smallskip}
    K (upper)&24&2015 Jun 18&$0.320 \pm 0.017$\\
\noalign{\smallskip}
\hline
\noalign{\smallskip}
  \end{tabular}
\label{tab:radiofluxes}
\begin{tablenotes}
\item[1]\textbf{$^{\rm a}$} Data retrieved from The VLA \textit{FIRST} Survey 
\end{tablenotes}
\end{table}

If we further assume that such a compact region is expanding at $\sim
c$, then the minimum energy in the synchrotron-emitting plasma would
imply jet kinetic powers $>3 \times 10^{44}$ erg s$^{-1}$,
which, although high, is substantially less than the radiative
luminosity of the quasar, and again this number would be reduced by
projection, or by assuming slower expansion speeds. It should be
noted, though, that the numerator of this calculation is the {\it
minimum} energy, and departures from equipartition, which would imply
higher energy densities in the synchrotron-emitting material, are very
common in larger-scale radio sources. Given these uncertainties, it
certainly does not seem impossible to imagine a scenario in which the
shock driven by the jet/lobe system responsible for the radio emission
gives rise to the fast bulk outflows detected in the X-ray and UV
spectra.
Although the shock driven by the jet provides the acceleration mechanism in this scenario, 
we would still expect the shocked gas to be photoionized by the central
continuum source, consistent with our photoionization model for the absorption.
\cite{Tombesi2014} have shown that both radio jets and UFOs can co-exist in
AGN, and may even be related.
Testing such a model would require very long baseline interferometric (VLBI) imaging sensitive
enough to detect and resolve the radio emission at mas resolution.

\section{Discussion}
\label{pg1211:discussion}

{\it Our simultaneous {\it Chandra} and {\it HST} observations are the first
definitive confirmation of an ultra-fast outflow detected simultaneously
in both X-ray and UV spectra.}
Highly ionized gas at an outflow velocity of
$-17\,300~\rm km~s^{-1}$ ($-0.0577c$) in our {\it Chandra} spectrum
is an excellent match to the broad Ly$\alpha$ absorption at
$-16\,980~\rm km~s^{-1}$ ($-0.0551c$) in our {\it HST}-COS spectrum
\citep{Kriss2017}.
Previous X-ray observations of PG\,1211+143 found evidence for outflows
clustered near several different velocities. In the original observations
using {\it XMM-Newton}, \cite{Pounds2003} identified an ultra-high velocity
outflow at $\sim -0.09c$  ($-27\,000~\rm km\,s^{-1}$),
detected only in very highly ionized gas
producing the Fe\,{\sc xxvi} K$\alpha$ transition.
Re-analysis of this same data set reaffirmed the detection of
ultra-high velocity gas, but at $-0.14 \pm 0.01~c$, with
additional high-ionization lines at $\sim-0.07 c$ \citep{Pounds2014}.
Most recently, the deepest observations to date of PG\,1211+143 using
{\it XMM-Newton} ($\sim$450 ks) in 2014 again detected ultra-high velocity
gas at $-0.129c$ and high-velocity gas at $-0.066 \pm 0.003 c$
\citep{Pounds2016b,Reeves2018}.
The UFO at  $-0.129~c$ is not seen in contemporaneous {\it NuSTAR} data
\citep{Zoghbi2015}, but a combined analysis of the {\it XMM-Newton} and
{\it NuSTAR} spectra show that the spectral structure around 7 keV is quite complex.
\cite{Pounds2016b} show that the $-0.129~c$ feature is quite variable,
both in column density and in ionization parameter.
Given the complexity of the spectrum around 7 keV and the lower column density
of the $-0.129 c$ feature in 2014, from a joint analysis of the
{\it NuSTAR} and {\it XMM-Newton} spectra, \cite{Lobban2016b} argue that it is
not surprising that it is not detectable in the {\it NuSTAR} data.

Subsequent analysis of the 2014 {\it XMM-Newton} RGS data by \cite{Reeves2018}
shows that two lower-ionization, lower velocity absorbers are
also present with velocities of
$-0.062 \pm 0.001 c$ ($v_{\rm out} = -18\,600 \pm 300$~km\,s$^{-1}$) and
$-0.059 \pm 0.002 c$ ($v_{\rm out} = -17\,700 \pm 600$~km\,s$^{-1}$), the latter of
which is a good match to the
$-17\,300$~km\,s$^{-1}$ ($-0.0577 c$) warm absorber we
detect in our {\it Chandra} spectrum.
However, the absorber we detect is slightly lower in ionization
(log $\xi = 2.87^{+0.32}_{-0.10}$ vs. log $\xi = 3.4 \pm 0.1$),
and nearly an order of magnitude lower in column density
($N_{\rm H} = 3 \times 10^{21}$~cm$^{-2}$ vs. $1 \times  10^{22}$~cm$^{-2}$). 
Both the velocity and the lower total column density are compatible with the
Ly$\alpha$ absorption detected in the simultaneous \hstcos spectrum
\citep{Kriss2017}.

{\it
Simultaneously detecting the same kinematic outflow with both
{\it Chandra} and {\it HST} provide the first opportunity to assess
the physical characteristics of an ultra-fast outflow using
both X-ray and UV spectra.
}
However, we note that this absorber is fairly high
ionization, both in the X-ray and the UV. This is consistent with our
detection of only broad Ly\,$\alpha$ in our UV spectrum--the ionization
is too high to produce significant populations of the usually seen UV ions.
\cite{Kriss2017} also do not detect UV absorption lines that might be
associated with a lower-ionization warm absorber, either in the COS spectrum
or in archival spectra from earlier epochs.
This is consistent with no evidence for a lower-ionization X-ray WA in
PG\,1211+143.
Despite the curvature in the {\it Chandra} spectrum that might suggest a
lower-ionization absorber, there are no detected absorption lines,
either in the X-ray nor in the UV.
The UV is especially sensitive in this regard. All X-ray WAs also show
UV absorption in C\,{\sc iv} \citep{Crenshaw2003} or O\,{\sc vi} \citep{Dunn2008}.
Although \cite{Tombesi2013} has suggested that WAs may be a lower-ionization
manifestation  of the same wind structure represented by UFOs, but at larger
distances from the black hole, this has been disputed by \cite{Laha2016}.
The lack of a low-ionization absorber in PG\,1211+143, unfortunately, does
not have much bearing on this dispute since $\sim 30$\% of sources
containing UFOs do not have associated WAs.

The lower ionization state of the gas in our {\it Chandra} observation is
expected, given the $\sim2 \times$ lower X-ray flux in 2015 compared to 2014.
Such an ionization response has been seen in longer, more extensive observations
of other UFOs.
The long {\it XMM-Newton} observation of IRAS~13224$-$3809 shows variability of
its high-ionization UFO in concert with variations in the X-ray flux
\citep{Parker2017b,Parker2017a}, consistent with the response of photoionized
gas.  However, we also see a significant decrease in total column density,
$\log N_{\rm H} = 21.5$~cm$^{-2}$ compared to $\log N_{\rm H} = 22.0$~cm$^{-2}$ \cite{Reeves2018} and $23.3$~cm$^{-2}$ \cite{Pounds2016a} in 2014.
Column density variations are also seen in other UFOs such as PDS~456, where
\cite{Reeves2016} suggest that the broad, variable soft X-ray absorption lines
they see are due to lower-velocity clumps in the overall outflow.
In PG\,1211+143 itself, \cite{Pounds2016a} find that both ionization and column
density vary in the 2014 {\it XMM-Newton} observation.

Although in \S\ref{pg1211:jet} we have presented a notional model for driving
the observed outflow in PG\,1211+143 via shocks from a jet associated with the
radio source, the most popular mechanism for explaining ultra-fast outflows is
via a wind driven from the accretion disk.
In both radiative and MHD models of accretion disk winds, the velocity of the
outflow is expected to reflect the orbital velocity at which the wind was
launched  \citep[e.g.,][]{Proga2000,Proga2003,Fukumura2010,Kazanas2012}.
Based on reverberation mapping, the black-hole mass of PG\,1211+143 is
$ 1.46 \times 10^8~M_{\odot}$ \citep{Peterson2004} (although this is poorly
constrained).
For an orbital velocity of $17\,300$~km\,s$^{-1}$, the wind would have
originated at $r = 6.5 \times 10^{15}$~cm, or $\sim 300$ gravitational
radii ($r_g$).
% Time to complete one orbit of the black hole is 270 days.
Interestingly, the half-light radius (at 2500 \AA) for an accretion disk
surrounding a black hole of this mass is approximately
$5 \times 10^{15}~\rm cm$ \citep[this estimate is based on the compilation of reverberation-mapping and
gravitational micro-lensing results in][]{Edelson2015}. 
At the rest UV wavelength of the observed Ly$\alpha$ absorption feature
$\rm (1240 \AA/1.0809 = 1147 \AA$), scaling by the $\lambda^{4/3}$ temperature
profile of typical accretion disks \citep{Edelson2015}, the half-light radius
of the UV continuum is then $1.8 \times 10^{15}~\rm cm$.
Given that the X-ray absorber fully covers the continuum source, and that the UV
absorber seems to cover only $\sim$40 percent, then the projected size of the outflow
is also roughly $1.8 \times 10^{15}~\rm cm$.
This suggests that the outflow originates only from a portion of the disk,
perhaps from selective active regions, or that it is restricted to a conical
volume with opening angle smaller than the inclination to our line of sight.
In the latter case, the outflow could obscure the far side of the disk
(and the full X-ray emitting region), but leave our line of sight to the near,
outer side of the disk unobstructed.

As shown by several authors \citep[e.g.,][]{King2010,Reeves2012,Nardini2015,King2015},
the high outflow velocity in UFOs and their often substantial column density
can lead to a large injection of energy into the interstellar medium of the
AGN host galaxy.
Even though our arguments above establish a plausible origin for the UFO
in the outer portion of the accretion disk
(at a few hundred gravitational radii),
assessing its mass flux and kinetic power depend on its overall extent and
covering fraction.
\cite{Kriss2017} discuss several alternatives for
determining the energy in the outflow observed in our joint campaign.
For the case in which the outflow is restricted to a thin shell near its
origin at the accretion disk, the impact is minimal. They find a minimum
mass flow of $> 0.013~M_{\odot}$~yr$^{-1}$,
and a minimum power in the outflow of $>1.2 \times 10^{43}$~erg\,s$^{-1}$.
Our SED for PG\,1211+143 (see \S5.1) gives a total bolometric luminosity
of $5.3 \times 10^{45}$~erg\,s$^{-1}$, so such an outflow would comprise
only 0.02\% of the total energy output of the AGN.
In contrast, feedback from AGN at levels of 0.5--5\% of their radiated
luminosity are required to have an evolutionary impact on the host
galaxy in most models \citep{DiMatteo2005,Hopkins2010}.
For the more likely case where the power in the outflow is comparable to
the minimum kinetic luminosity in the jet
(\S\ref{pg1211:jet}: $3 \times 10^{44}~$erg\,s$^{-1}$), the outflow and the
jet would be injecting mechanical energy at 0.6\% of the AGN radiated
luminosity, which is sufficient to have an impact.
A definitive answer to whether the feedback from this UFO affects the host
galaxy, however, requires a conclusive determination of its total size
and extent.

\section{Summary and Conclusions}
\label{pg1211:conclusions}

To summarize,
we observed the optically bright quasar \pg with the \chandra-\hetgs
for a total of 433\,ks in 2015 April as part of a program that
simultaneously took \hstcos and \jvla observations.  In this paper we
have used the \hetgs X-ray spectra averaged over 390\,ks, when the
source was at low brightness. We have utilized \xstar photoionization
modeling to probe the physical conditions in the ionized absorbers of
this quasar. We have compared the results of this analysis to the
findings from the \hstcos study \citep{Kriss2017}.
The key findings from both our {\it Chandra} and {\it HST} analysis are:

1.~The combined 1st-order \chandra-\meg and -\heg gratings spectra 
shows that the hard X-ray spectrum of \pg is well described by a
simple power law 
above $\sim 1$\,keV in the rest frame, and a soft excess below $\sim
1$\,keV.
We use a phenomenological model consisting of an absorbed Comptonized accretion
disk to characterize the shape of the soft X-ray continuum.

2.~We also have identified three emission lines in the hard X-ray
spectrum, which are consistent with the K-shell, He-like and H-like
iron lines at 6.41\,keV (Fe\,K$\alpha$), 6.67\,keV (Fe\,He$\alpha$),
and 6.96\,keV (Fe\,Ly$\alpha$)/7.05\,keV (Fe\,K$\beta$) in the rest
frame, respectively. We did not detect any clearly identifiable
blue-shifted Fe absorption lines, in contrast with the \xmm
observations \citep{Pounds2003}. This could be due to the low
signal-to-noise ratio of the \heg spectra at high energy.

3.~We discovered absorption lines from H-like and He-like ions of Ne, Mg, and Si; their observed wavelengths
are consistent with an outflow velocity of
$-$17\,300 km\,s$^{-1}$ ($z_{\rm out}\sim-0.0561$)
relative to systemic.
Their ionic column densities and turbulent velocities are not
the same for all ionic species, which are related to blending
and/or contamination from other atomic transition lines, as well as
insufficient counts.

4.~The absorption lines have been modeled using the photoionization
\xstar grid constrained by the ionizing SED constructed using the
simultaneous \jvla radio, \hstcos UV, and \chandra X-ray \hetgs data,
as well as by archival \hstfos UV and infrared measurements.  The
absorption lines from H-like and He-like ions of Ne, Mg, and Si were
best fitted with a highly ionized warm absorber with ionization
parameter of $\log \xi = 2.87$ erg\,s$^{-1}$\,cm, column density of
$\log N_{\rm H} = 21.47$ cm$^{-2}$, and 
an outflow velocity of
$-$17\,300 km\,s$^{-1}$ ($z_{\rm out}\sim-0.0561$).
This velocity is in reasonable
agreement with the component at $\sim -0.06c$
 ($-18\,000~\rm km\,s^{-1}$) 
measured
in the \xmm observations, albeit with dissimilar physical conditions
\citep [$\log \xi = 3.4$ erg\,s$^{-1}$\,cm, $\log N_{\rm H} = 22.0$--$23.3$
  cm$^{-2}$;][]{Pounds2016a,Reeves2018}. Moreover, we did not identify any
additional spectral features associated with higher velocity
components (e.g., the $\sim -0.129c$ [$-38\,700~\rm km\,s^{-1}$],
$\log \xi = 4$ erg\,s$^{-1}$\,cm,
and $\log N_{\rm H} = 23.57$ cm$^{-2}$ absorber measured by
\citealt{Pounds2016a}), which are due to extremely low signal-to-noise ratio
over those energy band.

5.~We have detected a broad blueshifted Ly$\alpha$ absorption line
that has a similar outflow velocity
($v_{\rm out} = -16\,980 \pm 10~\rm km~s^{-1}$, $z_{\rm out} = -0.0551$)
as the X-ray absorber, and could be its likely counterpart \citep{Kriss2017}. The apparent
optical depth of the Ly$\alpha$ absorption line profile yields an
\hi\ column density of $\log N_{\text{\hi}} \gtrsim 14.5$~cm$^{-2}$
(assuming $C_f = 1$). The ionization parameter ($\log \xi = 2.87$) of
our best-fit X-ray absorber corresponds to the \hi\ ionization
fraction of $2.99\times 10^{-8}$. From our best-fit total hydrogen
column density ($\log N_{\rm H} = 21.47$~cm$^{-2}$), we obtain $\log
N_{\text{\hi}} = 13.95_{ }^{+0.26}$~cm$^{-2}$, which is roughly
consistent with the empirical \hi\ column density of the Ly$\alpha$
line.

6.~While inconclusive, our VLA observations hint at a possible tantalizing 
scenario for VLBI observations to test, in which the shock driven by the 
jet/lobe system responsible for the radio emission may be connected to 
the X-ray and UV-detected bulk outflows.

Our \chandra-\hetgs and \hstcos
observations of \pg are the first evidence for the \emph{same}
ultra-fast outflow occuring in both X-ray and UV spectra.  Crucial
to this discovery were spectrometers with velocity resolutions
well-matched to the width of the absorption lines.  Verifying these
results, searching for the additional absorption systems suggested by
the \xmm spectra, and studying the variations of these absorbers with
X-ray flux and/or spectral shape will either require significantly
longer \chandra-\hetgs spectra, or a high resolution X-ray
spectrometer with significantly higher effective area.  For the latter
possibility, the \arcus mission, recently accepted for Phase A study,
would provide 2--3$\times$ the \heg resolution and $>20\times$ the
\heg+\meg effective area at energies $\approx0.2$--1\,keV \citep{Smith2016},
and would allow us to perform a systematic study of the ultra-fast
outflows in \pg.

\acknowledgments

We thank the anonymous referee for helpful comments and corrections. 
We gratefully acknowledge the financial support of the \textit{Chandra} X-ray Center (CXC) grant
GO5-16108X provided by the National Aeronautics and Space Administration (NASA). The CXC is operated by the Smithsonian Astrophysical Observatory (SAO) for and on behalf of the NASA under contract NAS8-03060. 
S.C. is supported by the SERB National Postdoctoral Fellowship (No.\,PDF/2017/000841). 
T.F. was partly supported by grant 11525312 from the National Science Foundation of China. 
J.N. acknowledges support from NASA through a Hubble Postdoctoral Fellowship (HST-HF2-51343.001-A). 
The computations in this paper were run on the Odyssey
cluster supported by the Harvard FAS Research Computing Group, and the
Stampede cluster at the Texas Advanced Computing Center supported by
NSF grant ACI-1134872. The eXtreme Science and Engineering Discovery Environment (XSEDE) computing resources were supported by National Science Foundation (NSF) grant ACI-1053575. 
The scientific results reported in this article are based on observations made by the \textit{Chandra} X-ray Observatory under the \textit{Chandra} proposal ID 16700515. 
This work
was supported by NASA through a grant for \textit{Hubble Space Telescope} (\textit{HST}) program number 13947
from the Space Telescope Science Institute (STScI), which is operated by the
Association of Universities for Research in Astronomy (AURA), Incorporated,
under NASA contract NAS5-26555. Support for Program 13947 was provided
by NASA through CXC grant GO5-16108X. 
This work is also based on observations made with the NASA/ESA \textit{HST}, 
obtained from the Hubble Legacy Archive. 
The National Radio Astronomy Observatory (NRAO) is a facility of the National
Science Foundation operated under cooperative agreement by Associated
Universities, Inc.

\begin{appendix}

For the fits discussed within this paper, we have used a combination
of binning by signal-to-noise and uniform channel binning.  Binning
was employed to allow the use of $\chi^2$ statistics, which are faster
to minimize (this is a non-trivial concern when using the
\warmabs model, which is an extremely slow to evaluate code). 
Furthermore, adding of data and binning of channels can serve to
average over systematic calibration uncertainties.  However, any
non-uniform binning, e.g., a signal-to-noise criterion, can introduce
biases in line fits.

For our data, the signal-to-noise per channel rapidly varies at
energies $< 1$\,keV, and there is no (small) set of uniform channel
binnings that achieves adequate signal-to-noise in these channels.
Thus, we choose mixed criteria (using the \isis \group
function) that ensures both a minimum signal-to-noise and a minimum
number of channels in the binned data.  In practice for these
particular data, \emph{only} the minimum channel criterion applies
above 1\,keV, as this is sufficient to ensure signal-to-noise $\ge 4$
in all of these channels.

\begin{figure*}
\begin{center}
\includegraphics[width=0.45\textwidth, viewport=35 155 575 750, clip, angle=0]{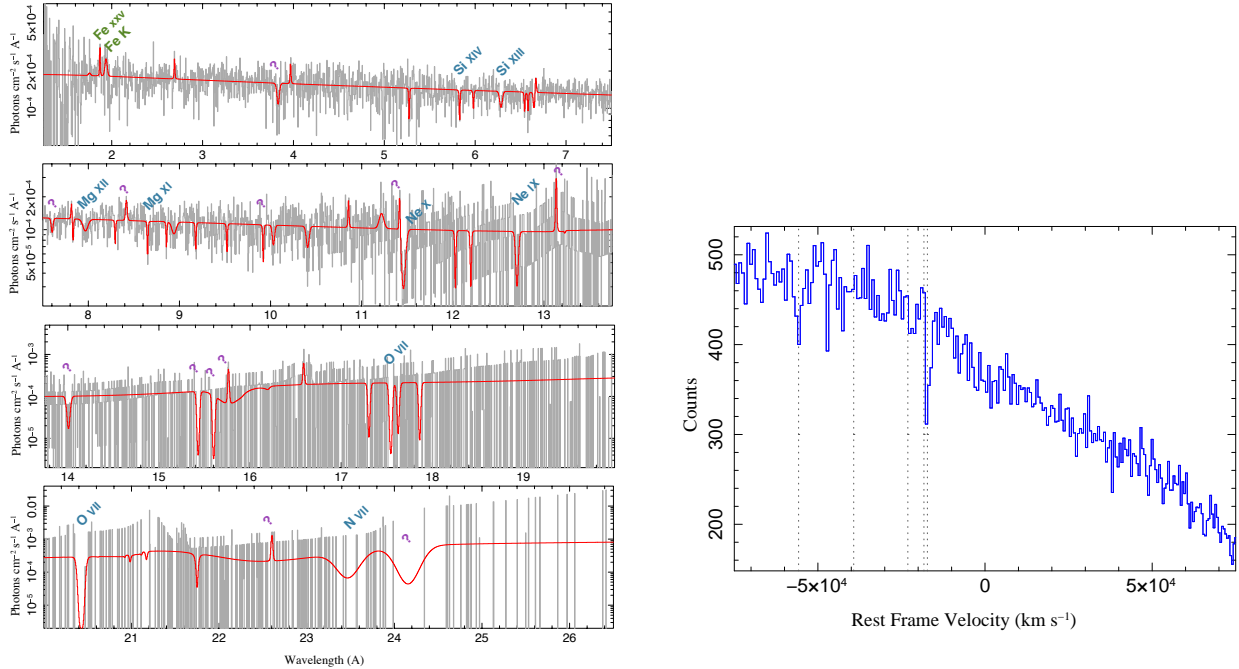}
\includegraphics[width=0.45\textwidth, angle=0]{figures/fig10_vshift.ps}
\end{center}
\caption{Left: Results of the blind line search, utilizing 51
  components.  The first 24 are presented in Table~\ref{tab:blind},
  along with possible line IDs, presuming an outflowing wind (in the
  cosmological rest frame of \pg) with 
  outflow velocity $v_{\rm out} = -17\,300~\rm km\,s^{-1}$ ($z_{\rm out}\sim -0.0561$). 
  Lines labeled with question marks are those found within the first
  24 components for which we have no suggested line ID.  Right:
  \chandra 1st-order detector counts stacked in velocity bins centered
  on the nine identified lines in Table~\ref{tab:blind}.  The dashed
  lines are the velocities of absorber components suggested by
  \citet{Pounds2016a}, as well as the 
$v_{\rm out} = -17\,300~\rm km\,s^{-1}$ ($z_{\rm out}=-0.0561$) absorber component 
  discussed in this work.}
 \label{fig:bsearch}
\end{figure*}

\begin{deluxetable*}{rrcrrrcrc}
\setlength{\tabcolsep}{0.03in}
\tabletypesize{\footnotesize}
\tablewidth{0pt}
\tablecaption{Results of Blind Line Search}
\tablehead{\colhead{$\lambda_{\rm obs}$} & \colhead{$\lambda_{\rm rest}$} &
  \colhead{$\sigma$} & \colhead{$\Delta C$} & \colhead{$-$EW} &
  \colhead{Component} & \colhead{ID} & 
  \colhead{$\lambda_{\rm lab}$} & \colhead{$z_{\rm out}$} \\
  \colhead{(\AA)} & \colhead{(\AA)} & \colhead{(\AA)} & & \colhead{(m\AA)} & & & \colhead{(\AA)} & (c)}
\startdata
  4.142 &  3.832 & 0.013 & $-$11.5 & $-$12.6 & 11 & \nodata & \nodata & \nodata \\
  6.173 &  5.830 & 0.002 & $-$13.1 &  $-$6.9 &  6 & Si{\sc xiv} (Ly$\alpha$)  & 6.182  & $-0.0586$ \\
  6.793 &  6.285 & 0.014 & $-$12.4 & $-$10.9 &  4 & Si{\sc xiii} (He$\alpha$) & 6.648  & $-0.0565$ \\
  8.219 &  7.603 & 0.009 &  $-$7.0 &  $-$7.6 & 21 & \nodata & \nodata & \nodata \\
  8.613 &  7.968 & 0.026 & $-$10.9 & $-$16.8 & 10 & Mg{\sc xii} (Ly$\alpha$)  & 8.421  & $-0.0555$ \\
  8.913 &  8.418 & 0.008 & $-$13.1 & $-$12.8 &  3 & \nodata & \nodata & \nodata \\
  9.160 &  8.651 & 0.002 &  $-$9.8 &  $-$9.0 & 13 & Mg{\sc xi} (He$\alpha$)  & 9.169  & $-0.0576$ \\
 10.504 &  9.920 & 0.003 &  $-$7.2 &  $-$9.8 & 20 & \nodata & \nodata & \nodata \\
 12.089 & 11.417 & 0.003 &  $-$9.9 &  22.7 & 19 & \nodata & \nodata & \nodata \\
 12.133 & 11.459 & 0.022 & $-$30.0 & $-$55.8 &  0 & Ne{\sc x} (Ly$\alpha$)  & 12.133 & $-0.0576$ \\
 13.453 & 12.705 & 0.013 &  $-$8.3 & $-$31.7 & 16 & Ne{\sc ix} (He$\alpha$) & 13.447 & $-0.0565$ \\
 13.909 & 13.136 & 0.002 & $-$10.1 &  35.7 & 15 & \nodata & \nodata & \nodata \\
 15.140 & 14.007 & 0.017 & $-$13.6 & $-$50.4 &  5 & \nodata & \nodata & \nodata \\
 16.338 & 15.430 & 0.009 & $-$12.5 & $-$52.3 &  9 & \nodata & \nodata & \nodata \\
 16.518 & 15.600 & 0.009 & $-$12.5 & $-$50.0 &  8 & \nodata & \nodata & \nodata \\
 16.691 & 15.763 & 0.002 &  $-$8.7 & 115.5 & 17 & \nodata & \nodata & \nodata \\
 17.739 & 16.753 & 1.132 & $-$15.6 & 141.7 &  2 & \nodata & \nodata & \nodata \\
 18.963 & 17.544 & 0.010 & $-$10.1 & $-$61.2 & 12 & O{\sc vii} (He$\beta$)  & 18.627 & $-0.0597$ \\
 22.080 & 20.428 & 0.025 &  $-$7.5 &$-$131.6 & 23 & O{\sc vii} (He$\alpha$) & 21.602 & $-0.0555$ \\
 23.933 & 22.603 & 0.004 &  $-$8.2 & 117.5 & 22 & \nodata & \nodata & \nodata \\
 24.431 & 23.073 & 0.600 & $-$22.4 & $-$50.5 &  1 & \nodata & \nodata & \nodata \\
 24.855 & 23.473 & 0.129 & $-$10.8 &$-$354.3 & 14 & N{\sc vii} (Ly$\alpha$) & 24.781 & $-0.0544$ \\
 25.584 & 24.161 & 0.130 &  $-$7.7 &$-$437.8 & 18 & \nodata & \nodata & \nodata \\
 29.610 & 27.964 & 0.099 & $-$13.2 &2631.8 &  7 & \nodata & \nodata & \nodata 
\enddata
\tablecomments{Results from a blind search to the unbinned, combined
  data, using a model consisting of an absorbed disk plus powerlaw
  with an exponential rollover.  An Fe line complex, consisting of Fe
K$\alpha$/K$\beta$, Fe{\sc xxv}, and Fe{\sc xxvi} was included in all
fits.  The columns give: the observed wavelength and width of the line
($\lambda_{obs}$, $\sigma$), the wavelength in the \pg
cosmological rest frame, the change in Cash statistic when including
the line, the line equivalent width (negative values refer to
absorption), the order in which the lines were added (Component
numbers 0--23), a suggested line ID, and the implied redshift in the
rest frame of \pg\ if this ID is correct.}\label{tab:blind}
\end{deluxetable*}

Our fits indicating an outflowing wind with a velocity in the \pg frame
of $-$17\,300 km\,s$^{-1}$ ($z_{\rm out}\sim-0.0561$)
are primarily being driven by the $\alpha$ lines of Helium-like and
Hydrogen-like Ne, Mg, and Si. To ensure that these lines are not being
biased by our binning (although five of the six primary lines in our
fits are within the uniformly binned portion of the spectrum at
energies $>1$\,keV), we have also performed a ``blind line search'' of
the data.  We grid the \heg data to the \meg bins, and combine all
spectra, but otherwise do not perform any further channel binning.  We
use the same continuum model as discussed above, namely an absorbed
disk plus powerlaw spectrum with exponential cutoff, and include line
emission from the Fe region.  We then loop through the spectra, adding
one line at a time which is allowed to freely range between emission
and absorption.  The initial line fit is constrained to a narrow range of
wavelengths (16 \meg channels, i.e., $\approx 0.18$\AA), but all
possible wavelength bins are searched.  The line with the greatest
change in fit statistic is retained.  After each step, all continuum
and line parameters are refit (within the constraints of the existing
wavelength region of the added line).  This process is repeated (51
times in Figure~\ref{fig:bsearch}, with the first 24 found lines listed
in Table~\ref{tab:blind}).  As the goal is to identify candidate
lines, we do \emph{not} calculate confidence intervals for the final
lines.

The putative Ne\,{\sc x} line is our single most significant residual,
with the remaining 5 lines from H-like and He-like Ne, Mg,
and Si all falling within the 17 most significant residuals.
Additionally, there are three other residuals that might be associated
with a $z_{\rm out} \approx -0.0561$ outflow.
These, however, fall within a very
low signal-to-noise region of the spectrum.  Several residuals are
broad, and are undoubtedly modifying the continuum fit.  Several could
be spurious noise features.  The rest remain unidentified.  However,
this blind search highlights those features that are driving the
\xstar and \warmabs models to identify an outflowing
component in the rest frame of \pg.

As a further diagnostic of possible absorber components, we take the
1st-order \chandra-\hetgs counts, and using the 9 potential H- and
He-like lines identified in Table~\ref{tab:blind}, stack the data into
cosmological rest frame velocity bins.  The results of this stacking
are also shown in Figure~\ref{fig:bsearch}  (note that in this
procedure, not every bin is statistically independent from one
another, as counts can be reused for different ions).  The strong
feature at a velocity in the rest frame of \pg\ of
$-$17\,300 km\,s$^{-1}$ ($z_{\rm out}\sim-0.0561$)
is apparent.  We also indicate the velocities of the absorber components
suggested by the analysis of \citet{Pounds2016a}.  There is a feature
near the $\sim-0.19c$ ($-57\,000~\rm km\,s^{-1}$) 
velocity found by \citet{Pounds2016a}, but we
have not found any single line-like residual at such a velocity.  It
is possible, however, that such an absorption component, if real, only
manifests significantly in a stacked analysis.

\end{appendix}

%\bibliographystyle{apj}                       %% AASTeX
%\bibliography{references}

\end{document}